\documentclass[12pt]{article}
\pdfoutput=1
\usepackage[a4paper]{geometry}
\usepackage{jheppub, amsmath,amssymb,amsfonts,amsxtra, mathrsfs, makeidx,graphics,graphicx,amsthm,epsfig, bm,longtable,float, color,tikz,mathtools,xfrac,footnote,rotating, lscape, makecell, environ,mathtools, empheq}

 \usepackage[utf8]{inputenc}
\usepackage{indentfirst}

\usepackage{subfig}
\usepackage{multirow}
\usepackage{adjustbox}

\usepackage{amsthm}
\usepackage{hyperref}

\usepackage{amsmath}
\usepackage{amsfonts}
\usepackage{commath}

\usepackage[english]{babel}
\usepackage{setspace}

\usepackage{epsfig,amssymb} 
\usepackage{amsmath}

\usepackage{amscd,color}
\usepackage{amsmath,graphicx}
\usepackage{verbatim,booktabs}

\usepackage{latexsym}
\usepackage{amsmath,amsfonts,amssymb,amsthm}
\usepackage{amsmath,amsthm}

\newcommand{\beq}{\begin{equation}}
\newcommand{\eeq}{\end{equation}}
\newcommand{\bea}{\begin{eqnarray}}
\newcommand{\eea}{\end{eqnarray}}

\newcommand{\eq}{\begin{equation}}
\newcommand{\feq}{\end{equation}}
\newcommand{\eqn}{\begin{eqnarray}}
\newcommand{\feqn}{\end{eqnarray}}

\newcommand{\ma}[1]{\mbox{$\mathcal{#1}$}}

\title{
\begin{center}
$T^{1,1}$ truncation on the spindle
\end{center}
}

\author[a]{Antonio %AMARITO
Amariti}
\author[b]{, Nicolò %SIMONA
 Petri}
\author[a,c]{and Alessia %SIMONA
 Segati}
\affiliation[a]{INFN, Sezione di Milano, Via Celoria 16, I-20133 Milano, Italy}
\affiliation[b]{Department of Physics, Ben-Gurion University of the Negev, Be'er-Sheva 84105, Israel.}
\affiliation[c]{Dipartimento di Fisica, Universit\`a degli Studi di Milano, Via Celoria 16, I-20133 Milano, Italy}
\emailAdd{antonio.amariti@mi.infn.it,petri@post.bgu.ac.il,\\alessia.segati@mi.infn.it}

\abstract{We study the compactification of the $\mathcal{N}=2$ AdS$_5$ consistent truncation of the conifold, in presence of a Betti vector multiplet, on the spindle.
We derive the BPS equations and solve them at the poles, computing the  central charge for both the twist and the anti-twist class, turning on the magnetic charge
associated to the baryonic symmetry.
Then, in the anti-twist class, where there are choices of the quantized flux that  give origin to a positive central charge, we numerically solve the 
BPS equations interpolating between the poles of the spindle.
We conclude by comparing our results with the one obtained from the analysis of the dual field theory, finding an exact agreement.}

\begin{document}

\maketitle

\section{Introduction}

The  compactification of superconformal field theories on curved space provides a way of defining new interacting fixed points in lower dimensions.
When such compactification preserves some supersymmetry there is a high control of the lower dimensional physics, because some observables can be traced through anomaly inflow and/or localization.
For example in the case of even dimensional field theories compactified on complex manifolds
one can integrate the anomaly polynomial over the compact sub-space to obtain the anomalies of the lower dimensional field theory.
For other dimensionalities similar results have been proven, with the help of SUSY localization.
This provides a strong tool to claim the existence of a lower dimensional interacting fixed point.
In many cases the complex manifold does not preserve any supersymmetry and it is necessary to turn on 
opportune background fluxes to restore the cancellations of the spin connection in the fermionic variations.
This idea, generically referred to as (partial) topological twist 
\cite{Witten:1988xj,Bershadsky:1995vm,Bershadsky:1995qy}, allows the analysis of large classes of models.
Furthermore the constructions discussed above have an interesting dual counterpart in the AdS/CFT correspondence. 
The original case was discussed in \cite{Maldacena:2000mw} 
for  branes wrapped on Riemann surfaces. This mechanism, denoted as flow across dimensions, was further studied  in \cite{Benini:2013cda}  and associated  to the dual c-extremization principle formulated in \cite{Benini:2012cz}.

The parallel treatment of the topological twist and of the flows across dimensions has been checked in many examples, and it provides a further check of the existence of the new superconformal fixed point in the lower 
dimensional theory.

Recently a new type of compactification has been considered \cite{Ferrero:2020laf}. The starting point consists of considering the compact space as an orbifold, instead of a manifold. If this orbifold is a spindle, topologically a two sphere with deficit angles at the poles, it is possible to show that some supersymmetry of the higher dimensional theory is preserved in an unusual way.
The Killing spinors are indeed not constant and chiral on the spindle.
It has been shown that there are two possible ways to preserve supersymmetry by turning on background  fluxes for the R-symmetry on the spindle. These two ways have been denoted as the twist and the anti-twist \cite{Ferrero:2021etw}.

In the case of the twist the integrated R-symmetry flux on the spindle corresponds to the Euler number of the spindle. This is the reason for using  the name twist for such a case, indeed  this property is shared by the common partial topological twist as well. Furthermore there is second possibility, denoted as the anti-twist, that does not have any counterpart in the standard partial topological twist.

The generality of such discussion allowed the study of this compactification in many setups, in various dimensions, showing very non trivial holographic matchings, and allowing to conjecture the existence of vast new families of SCFTs 
\cite{Ferrero:2020twa,Hosseini:2021fge,Boido:2021szx,Ferrero:2021wvk,Ferrero:2021ovq,Couzens:2021rlk,Faedo:2021nub,Ferrero:2021etw,Giri:2021xta,Couzens:2021cpk,Karndumri:2022wpu,Cheung:2022wpg,Cheung:2022ilc,Couzens:2022yjl,Suh:2022olh,Couzens:2022yiv,Couzens:2022aki,Couzens:2022lvg,Faedo:2022rqx,Boido:2022mbe,Suh:2022pkg,Inglese:2023wky,Suh:2023xse}.

In 4d SCFTs many predictions have been further made by the field theory analysis corresponding to the integration of the anomaly polynomial. Many of these predictions have been holographically checked, both from a 10 or 11d perspective, from the analysis of consistent truncations in gauged supergravity.

In this last case for example it has been possible to check the behaviour of the universal twist and anti-twist and  
in general the study of the $U(1)^3$ STU gauged supergravity has allowed to match the expected result for
the case of $\mathcal{N}=4$ SYM. 
Very recently a truncation with hypermultiplets has been considered as well. It corresponds to the Leigh Strassler
$\mathcal{N}=1^*$ fixed point and it has been shown that also in this case the expected dual results can be reproduced from  the supergravity dual description.
The case of the topological twist in this case was studied from 
the supergravity perspective in  \cite{Bobev:2014jva}.

One of the most remarkable results of \cite{Arav:2022lzo} was that the central charge of the theory compactified on the spindle can be obtained without the knowledge of the full solution of the BPS equation. It has been shown indeed that the correct central charge can be obtained by solving these equations only at the poles of the spindle, i.e. by  specifying the 
boundary conditions on the fields and the conserved magnetic charges  in terms of the data at the poles of the spindle.
This analysis at the poles is also a necessary step for constructing the numerical solution with the flavor twist because it fixes most of the boundary conditions when solving the BPS equation.

Motivated by the results of   \cite{Arav:2022lzo} here we study another 5d $\mathcal{N}=2$ consistent truncation with two vector multiplets and two hypermultiplets originally found in \cite{Cassani:2010na,Halmagyi:2011yd}. This truncation is associated to the Klebanov Witten theory and, due to the Higgs mechanism triggered by a scalar in a hypermultiplet, one vector field becomes  massive. The two remaining massless vector fields are the graviphoton and the  so called Betti vector. This structure of massless vector fields allows non-trivial comparisons with the field theory results
in terms of  the magnetic fluxes for the R-symmetry and  the  baryonic symmetry.
The role of baryonic symmetries in the case of the topological twist  was then exploited in \cite{Benini:2015bwz,Amariti:2016mnz}. 

Here we study the compactification of this model on the spindle, along the lines of the analsys of  \cite{Arav:2022lzo}.
We find that also in this case the central charge can be extracted simply from the pole data and then we solve numerically the BPS equation in order to construct the full AdS$_3$ solution.
As a consistency check we also show that our results are  in agreement with the ones expected from the dual field theory for the anti-twist class.

The paper is organized as follows.
In section  \ref{sec:5dtrunc} we review the 5d setup corresponding to the 5d $\mathcal{N}=2$ Betti vector  truncation found in \cite{Cassani:2010na} for the conifold. 
In section \ref{sec:3dansatz}  we study the BPS equations and the Maxwell equations for the AdS$_3$ ansatz on the spindle, turning on suitable magnetic fluxes for the gauge fields.
Then in section \ref{sec:ccpoles} we solve these equations  at the poles of the spindle. This analysis fixes the boundary conditions for 
many of the scalars and it imposes the necessary constraints on the fluxes. Then we show that these solutions are enough to 
compute the central charge from the Brown-Henneaux formula.
In section \ref{sec:BPSsol} we provide a complete solution of the BPS equations. First we turn-off the magnetic charge for the baryonic symmetry reducing to minimal supergravity. In this case we recover the analytic results of \cite{Ferrero:2020twa} for the universal anti-twist
and we match it with the result obtained in section \ref{sec:ccpoles} from the analysis of the BPS equations at the poles of the spindle.
Then we provide the numerical solution of the BPS equations in presence of non-vanishing baryonic magnetic charge, again finding an agreement with the result obtained from the pole data.
In section \ref{sec:FT} we then compare our findings with the calculation of the central charge for the conifold obtained from the dual field theory analysis. In this case we match the result by turning off the magnetic charges associated to the mesonic symmetries of the dual field theories, that are indeed invisible in the 5d truncation considered here.
In section \ref{sec:conc} we conclude, speculating over possible future directions.
We also add appendix \ref{generalSUGRA} to summarize the main features of $\ma N=2$ matter-coupled supergravity in five dimensions and appendix \ref{sec:app1} where we discuss further details of the quaternionic geometry of the specific 5d supergravity model we consider in this paper.

\section{The supergravity set-up}
\label{sec:5dtrunc}

In this section we introduce the $\ma N=2$ supergravity set-up in five dimensions. We firstly provide a short summary on $\ma N=2$ consistent truncations of Type IIB supergravity on squashed Sasaki-Einstein manifolds and we then focus on compactifications defined by the conifold as the internal manifold. In this regards, we introduce the $\ma N=2$ 5d gauged supergravity model associated to this truncation and we study its scalar manifold with particular focus on the gauging. Finally we discuss a further truncation of fields in 5d restricting to those moduli which capture the physics of AdS$_3\times \Sigma$ backgrounds where $\Sigma$ is the spindle.

\subsection{Type IIB on the conifold and $\ma N=2$ supergravity}\label{conifold5dsugra}

Let's start with a (very) brief summary on $\ma N=2$ consistent truncations of Type IIB supergravity over the (squashed) conifold
\begin{equation}
 T^{1,1}=\frac{SU(2)\times SU(2)}{U(1)}\,.
\end{equation}
 Such compactifications belong to a general class of consistent truncations on 5d squashed Sasaki-Einstein manifolds, which have been extensively studied in the literature (see for instance \cite{Acharya:1998db,Gauntlett:2010vu,Cassani:2010uw,Cassani:2010na,Liu:2010sa,Skenderis:2010vz,Bena:2010pr,Liu:2011dw,Halmagyi:2011yd}).
More specifically in \cite{Gauntlett:2010vu,Cassani:2010uw} reductions over squashed Sasaki-Einstein manifolds to the $\ma N=4$ 5d universal sector were constructed, then in \cite{Cassani:2010na,Bena:2010pr} this class of truncations was extended to the non-trivial second cohomology forms on $T^{1,1}$. The resulting lower-dimensional theory is a $\ma N=4$ 5d gauged supergravity coupled to two vector multiplets, coming from the universal sector, plus a third vector multiplet. The latter is called the Betti multiplet and it is associated to left-invariant modes acting on the conifold.

The bosonic field content of this $\ma N=4$ supergravity includes the 5d gravitational field, the graviphoton $A^0_\mu$, 8 vectors and 16 real scalar fields. We need now to impose a further truncation to select the $\ma N=2$ sector.
As it was showed in \cite{Cassani:2010na}, truncating to the $\ma N=2$ sector is not trivial since it requires the truncation either of the $\ma N=2$ Betti-vector multiplet or of the Betti-hypermultiplet. In this way one obtains two inequivalent theories. 

We will focus on the $\ma N=2$ 5d supergravity retaining in its spectrum the Betti vector\footnote{The ``twin" $\ma N=2$ theory is defined by truncating away the Betti-vector \cite{Cassani:2010na}. The matter content of this theory is featured by one vector multiplets and three hypermultiplets. In this case the coset manifold is given by $\ma M=SO(1,1)\times \frac{SO(4,3)}{SO(4)\times SO(3)}$.}. Such a theory is described by the coupling to two vector multiplets and two hypermultiplets whose scalar geometry is encoded in the following manifold \cite{Cassani:2010na},
\begin{equation}\label{model}
 \ma M=SO(1,1)^2\times \frac{SO(4,2)}{SO(4)\times SO(2)}\,.
\end{equation}
Let's thus explore with more detail the supergravity model defined by the scalar manifold \eqref{model}. To this aim we will follow the notation of \cite{Louis:2016msm}.
The $\ma N=2$ matter multiplets include 10 scalar fields and 3 vector fields
\begin{equation}\label{fieldcontent}
 \{u_1,u_2,u_3,k,a,\phi,\,b^i,\bar b^i  \}\quad \text{with}\quad i=1,2 \qquad \text{and} \qquad  A^I_\mu\quad \text{with}\quad I=0,1,2\,.
\end{equation}
Apart from the graviphoton $A^0_\mu$, the above fields are organized into two vector multiplets, defined by the real scalars $\{u_2,\,u_3\}$ and the two vectors $A^x_\mu$ with $x=1,2$, and two hypermultiplets, parametrized by the scalars $\{u_1,k,a,\,\phi,b^i,\bar b^i \}$ where $b^1$, $b^2$ are written in complex notation. 

We first analyze the vector multiplet sector. The two real scalars parametrize the Very Special Real manifold $SO(1,1)^2$. We can organize the moduli fields following the general analysis of appendix \ref{generalSUGRA},
\begin{equation}
 \phi^x = (u_2, u_3 ) \qquad\text{and} \qquad  g_{xy} =
  \begin{pmatrix}
    4 & 0 \\
    0 & 12
  \end{pmatrix}\,,
\end{equation}
where $g_{xy}$ is metric on the manifold $SO(1,1)^2$.
As explained in appendix \ref{generalSUGRA}, a Very Special Real manifold can be defined through the embedding relation $ C_{IJK} h^Ih^Jh^K = 1$ where $h^I(\phi)$ are homogeneous coordinates. For our model these are given by
\begin{equation}
  h^0 = e^{4u_3}, \qquad h^1 = e^{2 u_2 - u_3}, \qquad h^2 = e^{-2 u_2 -2 u_3}\,,
\end{equation}
 where the symmetric tensor $C_{IJK}$ has a unique non-vanishing component given by $C_{012}=1/6$. Through the general relation written in \eqref{SKa} we can thus derive the metric $a_{IJ}$ on the embedding manifold. Such quantity defines the coupling in the action of vector-scalars with gauge fields,
 \begin{equation}
\qquad a_{IJ} =
  \begin{pmatrix}
    \frac{1}{3} e^{-8u_3} & 0 & 0 \\
    0 & \frac{1}{3} e^{- 4u_2 + 4u_3} & 0 \\
    0 & 0 & \frac{1}{3} e^{4u_2 + 4u_3}
  \end{pmatrix}
  .
\end{equation}

Let's consider now the hypermultiplet sector. This is defined by the quaternionic manifold $\frac{SO(4,2)}{SO(4)\times SO(2)}$ which is spanned by the fields $\{u_1,k,a,\,\phi,b^a,\bar b^a \}$. We point out that the scalars $a$ and $\phi$ can be also written in complex notation as $\tau=a+ie^{-\phi}$, where the complex modulus $\tau$ results directly from the reduction of the axio-dilaton of Type IIB supergravity. As discussed in appendix \ref{generalSUGRA}, the scalars parametrizing a quaternionic manifold are organized in quadruples $q^X$. Following the notation of \cite{Halmagyi:2011yd, Louis:2016msm} we can write for our model $q^X = (u_1, k, a, \phi,\, b^1,\,\bar{b}^1,\, b^2, \,\bar{b}^2 )$. Then the line element takes the following form
\begin{equation}
  \begin{split}
    g_{XY}  dq^X dq^Y = & -2 e^{-4u_1} M_{ij} \bigl( b^i d\bar{b}^j + \bar{b}^i d b^j \bigr) - 4 du_1^2  - \frac{1}{4} d\phi^2 - \frac{1}{4} e^{2\phi} da^2  \\
    & - \frac{1}{4} e^{-8u_1} \Bigl[ dk + 2 \varepsilon_{ij} \bigl( b^i d\bar{b}^j + \bar{b}^i db^j  \bigr)  \Bigr]^2 ,
  \end{split}  
\end{equation}
where
\begin{equation}
  M_{ij} = e^\phi 
  \begin{pmatrix}
    a^2 + e^{-2\phi} & \quad -a \\
    -a & 1
  \end{pmatrix}
  .
\end{equation}
We point out that the matrix $M_{ij}$ is covariant under $SL(2,\mathbb{R})$ symmetry inherited from Type IIB supergravity. This scalar geometry was already studied in \cite{Halmagyi:2011yd, Louis:2016msm,DallAgata:2021nnr}. We refer to appendix \ref{sec:app1} for the derivation of quaternionic structures $\vec J$ and the $SU(2)$ spin connections $\vec \omega$.

On the quaternionic manifold we have gauged symmetries defined by the following set of abelian Killing vectors $k_I=k_I^X\partial_X$,
\begin{equation}\label{Kvectors}
  \begin{split}
    k_0 & = -3ib^1 \partial_{b^1} - 3ib^2 \partial_{b^2} + 3i\bar{b}^1 \partial_{\bar{b}^1} + 3i\bar{b}^2 \partial_{\bar{b}^2} - Q \partial_k,\\
    k_1 & = 2 \partial_k, \\
    k_2 & = 2 \partial_k,
  \end{split}
\end{equation}
where $Q$ is a constant.
Given the Killing vectors on the quaternionic manifold we can also introduce the associated Killing prepotentials $P^r_I$,
\begin{equation}
  \begin{split}\label{momentMaps}
    P_0^r & =
    \begin{pmatrix}
      \frac{3}{2} e^{-2u_1-\frac{\phi}{2}} \bigl( b^1 -i a e^\phi ( b^1 - \bar{b}^1 ) +  i e^\phi ( b^2 -\bar{b}^2 ) + \bar{b}^1 \bigr) \\
     \frac{3}{2} e^{-2u_1-\frac{\phi}{2}} \bigl( i b^1 + a e^\phi (b^1 + \bar{b}^1) - e^\phi ( b^2 + \bar{b}^2 ) - i \bar{b}^1 \bigr) \\
      -\frac{3}{2} + e^{-4u_1} \bigl( \frac{Q}{4} - 3 i (b^1 \bar{b}^2 - b^2 \bar{b}^1 ) \bigr)
    \end{pmatrix}
    , \\
    P_1^r & =
    \begin{pmatrix}
      0 \\
      0 \\
      - \frac{1}{2} e^{-4u_1}
    \end{pmatrix}
    , \qquad P_2^r =
    \begin{pmatrix}
      0 \\
      0 \\
      - \frac{1}{2} e^{-4u_1}
    \end{pmatrix}
    .
  \end{split}
\end{equation}
The full bosonic Lagrangian of this model is given in \cite{Halmagyi:2011yd, Louis:2016msm}. Specifically, the scalar potential can be obtained by specifying the general expression \eqref{scalarpotential} with the datas on the scalar geometry given in this section. We point out that the Killing vectors \eqref{Kvectors} satisfy abelian commutation relations and generate the gauge group $U(1)^2\times \mathbb{R}$. After such a gauging the scalars turn out to be charged under the subgroup $U(1)\times\mathbb{R}$ through the gauge vectors $A^0_\mu$ and $QA_\mu^0-2A_\mu^1-2A_\mu^2$ \cite{Halmagyi:2011yd, Louis:2016msm}.

\subsection{The model}

In this section we specify us to a further truncation of the 5d supergravity model introduced in \ref{conifold5dsugra}. Our aim is to retain the minimum set of fields needed to capture the oscillations of Type IIB supergravity described by a warped product of an AdS$_3$ factor with the spindle. In this regards we may firstly truncate the fields
\begin{equation}
a=0\qquad \text{and}\qquad \phi=0\,.
\end{equation}
This condition is equivalent to restrict to Type IIB systems with trivial axio-dilaton profile. Then we can also impose that
\begin{equation}
 b^1=\bar b^1=b^2=\bar b^2=0\,,
\end{equation}
which is equivalent to exclude those scalar fields associated to 3-form fluxes in Type IIB, namely 5-brane contributions. 
It follows that with this truncation we focus only on 3-brane systems.
Summarizing we look at those solutions featured only by the hyperscalar $u_1$, the vector multiplet-scalars $u_2$, $u_3$ and the three vectors $A_\mu^I$. The remaining scalar $k$ is a flat direction of the potential.

Given the above truncation, the Killing vectors \eqref{Kvectors} boil down to $ k_0 =  - Q \,\partial_k,\,\,\, k_1  = 2 \partial_k,\, \,\,k_2  = 2 \partial_k$. From this expression one can observe that the scalar $k$ gets charged under the vector $QA_\mu^0-2A_\mu^1-2A_\mu^2$, which in turns becomes massive. As far as the moment maps \eqref{momentMaps} are concerned, only the $r=3$ $SU(2)$-components survive, leading to
\begin{equation}
  \begin{split}\label{momentMapsTruncated}
    P_0'^3 & =-\frac{3}{2} + \frac{Q}{4}\,e^{-4u_1}
    , \qquad P_1^3=P_2^3 = - \frac{1}{2} e^{-4u_1}
    .
  \end{split}
\end{equation}
For such supergravity model we can introduce a superpotential as it follows,
\begin{equation}
  \label{WW}
  W = \sqrt{\frac{2}{3}} h^I P_I^3 = \sqrt{\frac{2}{3}}\, \biggl( \frac{1}{4} e^{-4u_1 - 2u_3} \bigl( Q e^{6u_3} - 4 \cosh(2u_2) \bigr) -\frac{3}{2} e^{4u_3} \biggr).
\end{equation}
The scalar potential can be thus derived by the general formula $V=2g^{\Lambda\Sigma}\partial_\Lambda W\partial_\Sigma W-\frac43\,W^2$ \cite{Halmagyi:2011yd} where $\Lambda, \Sigma$ include vector multiplet and hypermultiplet fields. This truncation contains the following AdS$_5$ vacuum for $u_{1,2,3}$,
\begin{equation}
\label{5dvacuum}
u_1=-\frac{1}{4} \log \frac{4}{Q} ,\quad u_2 = 0,\quad u_3=\frac{1}{6} \log \frac{4}{Q}
\end{equation}
with all the other scalars in the hypermultiplets have been set to zero.

\section{AdS$_3\times \Sigma$ geometry and  BPS equations}
\label{sec:3dansatz}

In this section we introduce the AdS$_3\times \Sigma$ Ansatz and we present the corresponding BPS equations (for details on the derivation of BPS equations see appendix \ref{app:BPS}). For the space $\Sigma$ we will take a compact spindle with conical singularities at the poles. Once presented the Ansatz and BPS equations we will also derive Maxwell equations for vector fields and study the corresponding conserved charge. In what follows we will adapt to the case of the conifold the analsys presented in \cite{Arav:2022lzo} on AdS$_3\times \Sigma$ geometries dual to Leigh-Strassler SCFT compactified on a spindle.

\subsection{The Ansatz and Maxwell equations}

Let's start with the following AdS$_3 \times \Sigma$ geometry \cite{Arav:2022lzo}
\begin{equation}
\label{ads3ansatz}
ds^2 = e^{2V(y)} ds^2_{\text{AdS}_3} + f(y)^2 dy^2 + h(y)^2 dz^2
\end{equation}
together with the gauge fields
\begin{equation}
\label{ansatz:A}
A^{(I)} = a^{(I)}(y) dz\,,
\end{equation}
where $ds^2_{\text{AdS}_3}$ is the metric of unitary AdS$_3$.
We suppose that the scalars $u_1, u_2, u_3$ are functions of $y$, while we take the hyperscalar $k$ linear along the $z$-direction, i.e. $k = \overline k z$. This prescription, originally given in \cite{Arav:2022lzo}, follows from Maxwell equations which imply that $k=k(z)$. Then in order to reproduce a set of Maxwell equations which are ODE along the $y$-direction (avoid terms as $\partial_zk$) we need that $k$ is linear.

The space $\Sigma$ in (\ref{ads3ansatz}) is a compact spindle, a weighted projective space $\mathbb{WCP}^1_{[n_N,n_S]}$ with conical deficit angles at the north and the south pole. 
The geometry is specified by the two co-prime integers $n_N \neq n_S$,  associated to the deficit angles 
$2 \pi \left(1-\frac{1}{n_{N,S}} \right)$ at the poles.
The azimuthal symmetry is parameterized by $\partial_z$, where the coordinate $z$ is periodic with period $\Delta z = 2 \pi$.
The coordinate $y$ is  compact, bounded by $y_N$ and $y_S$ (with $y_N < y_S$), i.e. finite values at the north and the south pole of the spindle.
It follows that function $h(y)$ vanishes at the poles and a crucial problem consists of finding the boundary conditions for the other fields at the poles of the spindle.

In order to study the Killing spinor equations and the equations of motion of the gauge fields it will be useful to work in the orthonormal frame 
\begin{equation}\label{flatbasis}
e^a = e^V \overline e^a,\qquad e^3 = f dy, \qquad e^4 = h dz, 
\end{equation}
where $ \overline e^a$ is an orthonormal frame for AdS$_3$. In this basis the field strengths takes the following form
\begin{equation}
f h F_{34}^{(I)} = \partial_y a^{(I)}.
\end{equation}
We can thus derive Maxwell equations specified to our Ansatz \eqref{ads3ansatz} and \eqref{ansatz:A}. We noticed that imposing that the scalars $u_1, u_2, u_3$ are defined over the $y$ direction and $k=\bar k z$, Maxwell equations can be easily integrated. Thus we can write them in the orthonormal frame as 
\begin{align}
  \label{gauge:eqns1}
   \frac{1}{3} e^{3V + 4u_3} \Bigl( e^{-4u_2} F^{(1)}_{34} - e^{4u_2} F^{(2)}_{34} \Bigr) & = \mathcal{E}_1,\\
     \label{gauge:eqns2}
   \frac{1}{3} e^{3V} \Bigl( e^{-8u_3} F^{(0)}_{34} + \frac{Q}{4} e^{4u_3} \bigl( e^{-4u_2} F^{(1)}_{34} +e^{4u_2} F^{(2)}_{34} \bigr) \Bigr) & = \mathcal{E}_2,\\
\partial_y \Bigl( \frac{1}{3} e^{3V-4u_2+4u_3} F^{(0)}_{34} \Bigr) & = - e^{3V-8u_1} f h^{-1} g D_zk,
 \end{align}
 where $\mathcal{E}_1$ and $\mathcal{E}_2$ are constants of motion and we defined $D_zk = \bar{k} - g Q a^{(0)} + 2 g a^{(1)} + 2 g a^{(2)}$. From the last equations we immediately notice that the scalar $k$ is charged under the vector $QA_z^0-2A_z^1-2A_z^2$, as we mentioned in previous section.
 
 \subsection{The BPS equations}
 
 In order to derive the BPS equations for the AdS$_3\times \Sigma$ geometry described in previous section we need to give a prescription on Killing spinors. Let's start by factorizing the spinor as it follows \cite{Arav:2022lzo},
\begin{equation}
  \epsilon = \psi \otimes \chi,
\end{equation}
where $\chi$ is a two-component spinor on the spindle and $\psi$ is a two-component spinor on AdS$_3$ satisfying
\begin{equation}
  \nabla_m \psi = - \frac{\kappa}{2} \, \Gamma_m \psi,
\end{equation}
where $\kappa = \pm 1$ specifies the two chiral cases with $\ma N=(2,0)$ or $\ma N=(0,2)$ supersymmetry. We outline the derivation of BPS equations in appendix \ref{app:BPS}. The Killing spinor analysis starts with the decomposition of the 5d gamma matrices that we choose as it follows
\begin{equation}
  \gamma^m = \Gamma^m \otimes \sigma^3, \qquad \gamma^3 = \mathbb{I}_2 \otimes \sigma^1, \qquad \gamma^4 = \mathbb{I}_2 \otimes \sigma^2,
\end{equation}
with $\Gamma^m = \bigl( -i \sigma^2, \sigma^3, \sigma^1 \bigr)$. From gravitino SUSY variations it turns out that the Killing spinor $\epsilon$ is defined in terms of a function $\xi(y)$ and it has the following form
\begin{equation}
\begin{split}\label{Kspinor}
 &\epsilon=\left[\cos\left(\frac{\xi}{2}\right)\mathbb{I}-\sin\left(\frac{\xi}{2}\right)\gamma^4 \right]\,\eta\qquad \text{with}\qquad \eta=e^{V/2}\,e^{isz}\,\eta_0\,,
 \end{split}
\end{equation}
where $s$ is a constant and $\eta_0$ is a constant spinor. The above structure of the Killing spinor characterizes the spindle geometry, for instance it was already obtained in \cite{Arav:2022lzo} for AdS$_3\times \Sigma$ geometries in Leigh-Strassler compactification.
We can thus write the BPS equations obtained by specifying the $\ma N=2$ SUSY variations of fermionic fields \eqref{susy:gr}, \eqref{susy:g} and \eqref{susy:h} to the AdS$_3$ Anstaz \eqref{ads3ansatz} and \eqref{ansatz:A} with a spinor $\epsilon$ of the form \eqref{Kspinor},
\begin{eqnarray}
\label{eq:resumeBPS}
 \xi' &=& 3 g   \, f\,W \cos \xi  +2 \kappa \, f\, e^{-V}\,, \nonumber \\
 V'&=& g\, f\,  W \sin \xi\,, \nonumber \\ 
  h' &=&h\, f\,\sin^{-1} \xi \bigl(2 \kappa  e^{-V} \cos \xi + g W (1+2 \cos ^2 \xi)\bigr) \,,\nonumber\\
  u_1' & =& \frac{3}{16} g \, f\,\partial_{u_1} W\,\sin^{-1}\xi\, ,\nonumber\\
  u_2'&=& -\frac{3}{4} g \, f\,  \partial_{ u_2}W \sin \xi \,, \nonumber \\
 u_3'&=&-\frac{1}{4}  g \, f\, \partial_{u_3}W \sin \xi\,, \nonumber \\
\end{eqnarray}
where $W$ is the superpotential defined in \eqref{WW}.
In addition to the first-order equations the analysis of SUSY variations leads to two algebraic constraints
\begin{align}
\label{constr}
 &(s-Q_z) = -\frac{h'}{2 f} \cos \xi + \frac{h}{\sqrt{6}} H_{34} \sin \xi, \\
\label{constrbis}
&\frac{3g}{2} \partial_{u_1}W \cos\xi  = h^{-1} \partial_{u_1} Q_z \sin\xi.
\end{align}
where $Q_\mu dx^{\mu}=Q_z\,dz$ is the connection associated to the supercovariant derivative $D_\mu\epsilon=(\nabla_\mu-i Q_\mu)\epsilon$ appearing in the gravitino variation \eqref{var:gr:2}. The tensor $H_{\mu\nu} \equiv h_I F^I_{\mu\nu}$ is introduced in \eqref{var:gr:2} and its non-zero components (in the flat basis \eqref{flatbasis}) are given by
\begin{equation}
H_{34}= 
\frac{1}{3} \left( e^{-4 u_3}  F_{34}^{(0)} + e^{2 u_3} \left(e^{-2 u_2} F_{34}^{(1)}+e^{2 u_2}F_{34}^{(2)}\right) \right) = - \sqrt6 \bigl( \kappa\, e^{-V}+ g\, W  \cos \xi \bigl) \,.
\end{equation}
Finally from the variations of gaugini and hyperini \eqref{BPSeqGaugino:1}, \eqref{BPSeqGaugino:2}  and \eqref{BPSeqHyperino} we can obtain the non-zero components of the field strengths $F^I_{34}$,
\begin{eqnarray}
\label{eqF34all}
e^{-4 u_3}  F_{34}^{(0)} &=& -\sqrt{\frac{3}{2}} g \cos \xi  \left(2 W +\partial_{u_3} W \right)- \sqrt6 \, \kappa  e^{-V} , \nonumber \\
e^{-2 u_2+2 u_3} F_{34}^{(1)} &=& -\frac{g}{2}\sqrt{\frac{3}{2}} \cos \xi  \left(4 W - \partial_{u_3} W + 3 \partial_{u_2} W  \right) - \sqrt6 \, \kappa  e^{-V}, \nonumber\\
e^{2 u_2+2 u_3}  F_{34}^{(2)} &=& -\frac{g}{2}\sqrt{\frac{3}{2}} \cos \xi  \left(4 W - \partial_{u_3} W - 3 \partial_{u_2} W  \right) - \sqrt6 \, \kappa  e^{-V}.\nonumber 
\end{eqnarray}

As it is discussed in \cite{Arav:2022lzo}, the analysis of BPS equation is simplified observing that we can integrate one out the first three differential equations in \eqref{eq:resumeBPS} obtaining
\begin{eqnarray}
\label{hintegrata}
h = \ell \, e^{V} \sin \xi
\end{eqnarray}
where $\ell $ is a constant that has to be determined.  
The BPS equation for $\xi'$ can be further simplified  by plugging in  \eqref{hintegrata}  and we obtain
\begin{eqnarray}
f^{-1} \xi ' = - 2 \ell^{-1} (s-Q_z) e^{-V}.
\end{eqnarray}
Similarly the constraint  (\ref{constr})  can be simplified to 
\begin{eqnarray}
\label{newconstr}
(s-Q_z)=-\frac{3}{2} g \ell e^V W  \cos \xi- \kappa  \ell.
\end{eqnarray}
Finally we can use the field strenghts \eqref{eqF34all} to get a fully explicit form of the conserved charges $\mathcal{E}_{1,2}$ obtained after having integrated out Maxwell equations in \eqref{gauge:eqns1} and \eqref{gauge:eqns2},
\begin{eqnarray}
\mathcal{E} _1&=& 3 g e^{3 V} \cos \xi -\frac{1}{\sqrt6} \kappa  e^{2 V-4 u_3} \left(Q e^{6 u_3} \cosh \left(2 u_2\right)+2\right),
\nonumber \\
\mathcal{E} _2&=&2 \sqrt{\frac23} \kappa  e^{2 \left(u_3+V\right)} \sinh \left(2 u_2\right)\,,
    \end{eqnarray}
    where we used the superpotential $W$ written in \eqref{WW}.
  
  We observe that with the redefinitions $g \to \frac{g}{3}, \,\,\, u_2 \to \beta, \,\,\, u_3 \to \alpha \,\,\, , u_1 \to 2 \sqrt 2 \varphi$ our equations take the form of (3.10) in \cite{Arav:2022lzo}
  %. However, we remind that the Type IIB interpretation of our 5d fields is completely different from that one of \cite{Arav:2022lzo}. {\color{red}  naive}

\section{Central charge from the pole data}
\label{sec:ccpoles}

In this section we compute the central charge of the dual field theory
obtained by the twist and the anti-twist of the truncation of $T^{1,1}$ without solving the BPS equations.
The relevant result is indeed that  this value can be predicted by specifying the boundary conditions 
at the poles of the spindle solution.
This does not guarantee the existence of the solutions, that requires to solve the BPS equation from the north to the south pole of the spindle and  that will be the subject of the analysis in section \ref{sec:BPSsol}. Nevertheless it is a notable result, already noticed in \cite{Arav:2022lzo} for the Leigh Strassler truncation.

\subsection{Simplifications at the poles}

At the poles $\ell \sin \xi \rightarrow 0$, then if $\ell \neq 0$ it follows that $\cos \xi_{N,S} = (-1)^{t_{N,S}}$ where $t_{N,S}=0$ or $t_{N,S}=1$. The poles are identified with $y_{N,S}$, where without loss of generality we can choose $y_N \leq y \leq y_S$. We have to impose also that 
\begin{equation}
| h' |_{N,S} = | \ell \sin' \xi |_{N,S} = \frac{1}{n_{N,S}}.
\end{equation}
This follows from the metric and from the assumption that the deficit angles at the poles are $2 \pi \left(1-\frac{1}{n_{N,S} }\right)$ where $n_{N,S}>1$.

From the BPS equation we observe the $\mathbb{Z}_2$ symmetry acting on $\{ h,a^{(i)},Q_z,s,\ell \}$.
This transformation leaves the frame invariant and it can be used to restrict the analysis to the region $h\geq 0$ and then, since $V \in \mathbb{R}$, also $\ell \sin \xi \geq 0$.

The combination $\ell \sin \xi $ is then positive and it vanishes at the two poles, with $y_N<y_S$. Its derivative is then positive at $y_N$ and negative at $y_S$. This can be formalized by introducing two further constants, $l_N=0$ and $l_S=1$ and requiring 
\begin{equation}
\ell \sin' \xi |_{N,S} = \frac{(-1)^{l_{N,S}}}{n_{N,S}}.
\end{equation}
The twist and the anti-twist are distinguished by the relation between the chiralities of the preserved spinors at the two poles, coincident and opposite respectively. Then, among the four choices of $(t_N,t_S)$, the cases $(0,0)$ and $(1,1)$ correspond to the twist and the other two options $(1,0)$ and $(0,1)$ correspond to the 
anti-twist.

Then the complete set of pole data that we have to specify correspond to $\{ l_{N,S},n_{N,S},t_{N,S}\}$.
The simplification occurred in (\ref{newconstr}) allows to express the quantity  $(s-Q_z)$ at the poles in term of these  data as
 \begin{equation}
 s-Q_Z |_{N,S} = \frac{1}{2n_{N,S}} (-1)^{t_{N,S}+l_{N,S} +1}.
  \end{equation}   
 By looking at the BPS equation obtained in formula (\ref{BPSeqHyperino})   we observe that it is necessary the $\partial_{u_1} W |_{N,S}=0$ at the poles, otherwise $u_1$ does not stay finite. A further consequence of this constraint, combined with  (\ref{newconstr}),  is that also $\partial_{u_1} Q_z |_{N,S}=0$.
  Another consequence is that the two reals scalars in the special geometry are constrained at the poles as
 $  Q e^{6 u_3}-4 \cosh (2 u_2) |_{N,S}=0 $.
 A further assumption (\emph{a posteriori} motivated by the numerical analysis in sub-section \ref{numal})
is that $u_1 |_{N,S} \neq 0$. It has been shown in \cite{Boido:2022mbe} that this assumption 
implies $D_z k |_{N,S} = 0$.

It is then useful to use these relations to re-consider the expressions obtained above for the conserved charges $\mathcal{E}_{1,2}$, using some of the simplifications occurred at the poles. 
By defining
\begin{eqnarray}
\label{defM12}
    M_{(1)}=g e^{4 u_3+V}, \quad M_{(2)}=-\kappa+3 \sqrt{\frac{3}{2}}  M_{(1)} \cos (\xi )
     \end{eqnarray}
the charges  can be written as 
\begin{eqnarray}
\label{epsilonpoles}
\mathcal{E}_1 &=&
\frac{M_{(1)}^2}{g^2}  \left( \sqrt{\frac{2}{3}}  M_{(2)} e^{-12 u_3}-\frac{\kappa  Q^2}{4 \sqrt 6}\right)+\frac{\kappa  Q}{4 \sqrt 6}  e^{2 \left(u_3+V\right)} \left(Q e^{6 u_3}-4 \cosh \left(2 u_2\right)\right)
\end{eqnarray}
and
\begin{eqnarray}
\label{epsilonpoles2}
\mathcal{E}_2^2 &=&\frac{ M_{(1)}^4}{6 g^4} \left(Q^2-16 e^{-12 u_3}\right)+\frac{ M_{(1)}^4}{6 g^4} \left(16 e^{-12 u_3} \cosh ^2\left(2 u_2\right)-Q^2\right)
\end{eqnarray}
and hence last terms in (\ref{epsilonpoles}) and (\ref{epsilonpoles2}) vanish at the poles.
From (\ref{newconstr}) and $W|_{N,S}=-  \sqrt{\frac{3}{2}}  e^{4 {u_3}_{N,S}}$ we also have

  \begin{eqnarray}
  M_{(1)}|_{N,S} =     \frac{\sqrt 6}{9} \,
\left(2 \kappa  (-1)^{-t_{N,S}}-\frac{(-1)^{l_{N,S}}}{\ell \,  n_{N,S}}\right)
  ,\quad
   M_{(2)}|_{N,S} = 
\kappa -\frac{(-1)^{l_{N,S}-t_{N,S}}}{\ell \, n_{N,S}}.
   \end{eqnarray}
 From the definition of $M_{(1)}$ in (\ref{defM12})  we see that requiring $u_3,V \in \mathbb{R}$ and $g>0$ imposes 
 that this $M_{(1)}$ is  positive. This reflects into the constraints $  M_{(1)}|_{N,S} >0$.

 The fact that $\mathcal{E}_{1,2}$ are constant, and then equal at the poles, implies the two following equations for $ u_3$ at the poles:
    \begin{eqnarray}  
  \left(
\begin{array}{cc}
-\frac{16}{Q^2} M^4_{(1)}|_N & \frac{16}{Q^2}  M^4_{(1)}|_S  \\
  M^2_{(1)}|_N  M_{(2)}|_N  & -  M^2_{(1)}|_S  M_{(2)}|_S  \\
\end{array}
\right) \cdot \left(
\begin{array}{c}
 e^{-12 u_{3N}} \\
 e^{-12 u_{3S}} \\
\end{array}
\right)
=
\left(
\begin{array}{c}
M^4_{(1)}|_S- M^4_{(1)}|_N  \\
\frac{\kappa  Q^2}{8} \left(  M^2_{(1)}|_N  -  M^2_{(1)}|_S  \right) \\
\end{array}
\right). \nonumber \\
     \end{eqnarray}
     
\subsection{Magnetic fluxes}     
Here we express the magnetic fluxes in terms of the pole data and of the fields evaluated at such poles.
This will allow us to express the constant $\ell$ in terms of the spindle data, without solving the BPS equations.
We start observing that  
\begin{equation}
F_{yz}^{(I)} = (a^{(I)})'  = (\mathcal{I}^{(I)})'  
\end{equation}
with \footnote{We hope that the notation $h^I$ for the sections does not generate confusion with respect to the scalar function $h(y)$ in (\ref{ads3ansatz}).}
\begin{eqnarray}
\mathcal{I}^{(I)} &=& \sqrt{\frac{3}{2}}\,  \ell \, e^V h^I
 \cos \xi  .
\end{eqnarray}
 This relation can be worked out by looking at the BPS equations studied above. A similar formula has been derived in \cite{Boido:2022mbe} by looking at the BPS equations for all the hyperscalars. Such a derivation is unnecessary even if it can be a useful check of validity of these relations.
  
 It follows that the (still not quantized!) fluxes can be expressed in terms of the pole data as
  \begin{equation}
  \label{fluxesquantsugra}
 \frac{p_I}{n_N n_S} = \frac{1}{2 \pi}   \int_{\Sigma} g F^{(I)} = g \mathcal{I}^{(I)} \big|_N^S
 \end{equation}
 with
  \begin{eqnarray}
&&
 \mathcal{I}^{(0)} |_{N,S}=\sqrt{\frac{3}{2}} \frac{\ell }{g} M_{(1)}|_{N,S} (-1)^{t_{N,S}},
 \\
 &&  \mathcal{I}^{(2)} |_{N,S}- \mathcal{I}^{(1)} |_{N,S}= \pm  \sqrt{\frac{3}{2}} 
\frac{ \ell (-1)^{t_{N,S}}  M_{(1)}|_{N,S} }{2 g} \sqrt{Q^2-16 e^{-12 u_{3N,S}}}
\end{eqnarray}
 where $\pm$ depends on the sign of $u_3$.  
 Observe that we are not claiming yet that the fluxes $p_{0,1,2}$ are correctly quantized and we will come back to this problem in a few, reading the correct normalization from the AdS/CFT correspondence.
 
In order to work with properly quantized charges we now fix $Q=4$ and consider the following (quantized) charges   \begin{eqnarray}
&&   p_R 
\equiv  \frac{1}{2}  ( p_0+p_1+ p_2) %\frac{1}{2 Q}  (Q p_1+4(p_2+ p_3))
=
\frac{1}{2}  \left(n_S (-1)^{t_N}+n_N (-1)^{t_S}\right),
\nonumber \\
&&
 p_F \equiv \frac{3}{4} (p_2- p_1) =
 \frac{ \text{sign}(u_3) g \,  n_N n_S }{2}   (\mathcal{I}^{(3)} |_N^S- \mathcal{I}^{(2)} |_N^S),
 \nonumber \\
&&
p_M \propto 2 p_0-p_1-p_2=0 %Q p_1-2p_2-2p_3=0
    \end{eqnarray}
where the coefficient of $p_F$ is chosen to match with the one of the baryonic symmetry in the holographic dual description. This coefficient can be extracted from the 't Hooft anomaly Tr$RB^2$ along the lines of the procedure discussed in \cite{Benini:2020gjh}. Requiring the quantization of this charge then imposes a constraint on the constant $\ell$ and no further constraints on the spindle data.
The charge $p_R$ is  quantized if $n_N (-1)^{t_S} + n_S (-1)^{t_N} \in 2 \mathbb{ Z}$.
The quantization of $p_F$ is obtained as follows. First we use the fact that $2p_0 = p_1+p_2$ that gives
the relation
$2 p_F=3 p_2-2 p_R$. This is a crucial relation to determine the quantization of $p_F$.
We indeed have that $p_R = \frac{3}{4} (p_1+p_2) \in \mathbb{Z}$ and we must also impose that 
 $p_F = \frac{3}{4} (p_1-p_2) \in \mathbb{Z}$. As anticipated above the proportionality coefficient  in this last relation comes from anomaly matching and it
 contains the informations on how the 5d solution is uplifted on $T^{1,1}$ to give a Type IIB solution, thanks to the AdS/CFT correspondence.
 Then the relations $p_R + p_F = \frac{3}{2} p_2  \in \mathbb Z$ and 
 $p_R - p_F = \frac{3}{2} p_1  \in \mathbb Z$  imply  $ \frac{3}{2} p_{1,2} \in \mathbb{Z}$. This tells us the correct normalization to impose in 
  (\ref{fluxesquantsugra}), i.e. the definition of the fluxes $p_{0,1,2}$  should be modified by multiplying it by the factor $3/2$, in order to have integer fluxes, say $\hat p_{0,1,2} \equiv \frac{3}{2} p_{0,1,2} \in \mathbb{Z}$.
Furthermore $p_R \pm p_F \in \mathbb{Z}$ tell us that
  $p_{R,F}$ must have the same parity.

 Then from (\ref{epsilonpoles2}) we have
   \begin{eqnarray}
 |\mathcal{E}_2| =  
% \frac{M_{(1)}^2|_{N,S}  }{\sqrt 6 g^2}  \sqrt{Q^2-16 e^{-12 u_3}}
 \frac{4 M_{(1)}^2|_{N,S}  }{\sqrt 6 g^2}  \sqrt{1- e^{-12 u_3  |_{N,S}}}
 \rightarrow
 \mathcal{I}^{(2)} |_{N,S} - \mathcal{I}^{(1)} |_{N,S} = \frac{3}{2} \frac{ g \ell}{ M_{(1)}|_{N,S}} |\mathcal{E}_2|  (-1)^{t_{N,S}}.\nonumber \\
     \end{eqnarray}
 From this relation  we have to require $ e^{-12 u_3 |_{N,S} } \in (0,1] $. This constraint becomes a constraint on the allowed values of the constant $\ell$ that will be computed below. 
  Then   using the fact that $\mathcal{E}_2$ is constant and equal at the poles we can write
  \begin{eqnarray}
 \mathcal{I}^{(2)}  \big|^S_{N} - \mathcal{I}^{(1)}  \big|^S_{N} 
 =
\frac{3}{2}
 g \ell   |\mathcal{E}_2|
 \left(
 \frac{(-1)^{t_{S}}}{M_{(1)}|_{S}} 
 -
 \frac{ (-1)^{t_{N}}}{M_{(1)}|_{N}} 
 \right).
 \end{eqnarray}
We can simplify this expression using the relation
\begin{eqnarray}
(-1)^{t_{S}} M_{(1)}|_{N}
 -
 (-1)^{t_{N}} {M_{(1)}|_{S}} 
  =(-1)^{t_S +t_N+1}  \sqrt \frac{2}{3}
 \frac{n_S (-1)^{t_N}+n_N (-1)^{t_S}}{ 3 \, \ell \,  n_N  \, n_S}
% \nonumber \\
  \end{eqnarray}
 where we used the fact that  the possible values taken by $t_S$ and $t_N$  are all the possible combinations of   $0$ and $1$.
  Then we arrive at
 \begin{eqnarray}
\frac{ p_F}{ g n_S n_N}
 \!=\!
  \frac{ 3}{4}(
 \mathcal{I}^{(2)}  \big|^S_{N} \!-\! \mathcal{I}^{(1)}  \big|^S_{N} )
 =
\frac{3}{4\sqrt 6}
 \frac{ g  |\mathcal{E}_2| (-1)^{t_S +t_N+1} }{  M_{(1)}|_{S} M_{(1)}|_{N}}
 \frac{n_S (-1)^{t_N}\!+\!n_N (-1)^{t_S}}{  n_N n_S}
 \end{eqnarray}
 while
 \begin{eqnarray}
 \frac{ p_R}{ g n_S n_N}=
 %  \frac{ 1}{2 Q} (Q \mathcal{I}^{(1)}  \big|^S_{N} + 4 \mathcal{I}^{(2)}  \big|^S_{N} +4 \mathcal{I}^{(3)}  \big|^S_{N} )
 \frac{ 1}{2} ( \mathcal{I}^{(0)}  \big|^S_{N} +  \mathcal{I}^{(1)}  \big|^S_{N} +\mathcal{I}^{(2)}  \big|^S_{N} )
=
\frac{1}{2 g} \left( 
 \frac{n_S (-1)^{t_N}+n_N (-1)^{t_S}}{ n_N n_S}
   \right).
%   \nonumber \\
   \end{eqnarray}
Comparing these last two expressions  we have

 \begin{eqnarray} 
 \label{eqell}
 \frac{p_F^2}{p_R^2}
 =
 \frac{ 3 g^4  \mathcal{E}_2^2 }{8 M_{(1)}^2|_{S} M_{(1)}^2|_{N}}
   \end{eqnarray}
   
 This equation allows to determine the constant $\ell$ in terms of the integers $n_{N,S}$, $t_{N,S}$, $l_{N,S}$ and $p_F$.  Solving (\ref{eqell}) for $\ell$ we obtain
     \begin{eqnarray} 
 \ell&=&
(-1)^{t_N+1}\frac{ n_N^4+n_N n_S^3+n_N^3 n_S+n_S^4-4 p_F^2 n_N n_S}{ \kappa  n_N n_S (n_N-n_S) (3(n_N+n_S)^2+4 p_F^2)}, \quad \text{for} \, (t_N,t_S) = (0,0) \, \text{or} \, (1,1) \nonumber \\
\end{eqnarray}
corresponding to the case of the twist, and 
    \begin{eqnarray} 
\ell&=&(-1)^{t_N}\frac{n_N^4-n_N n_S^3-n_N^3 n_S+n_S^4+4 p_F^2 n_N n_S}{ \kappa  n_N n_S (n_N+n_S) (3(n_S-n_N)^2+4 p_F^2)},  \quad \text{for} \, (t_N,t_S) = (1,0) \, \text{or} \, (0,1) \nonumber \\ \nonumber \\
 \end{eqnarray}
 corresponding to the case of the  anti-twist. 

\subsection{Central charge from the pole data}

We are now ready to compute the central charge from the pole data. These last correspond to the integers
$\{ l_{N,S},n_{N,S},t_{N,S}\}$ and in addition to the constant $p_F$. 
The central charge is obtained from the Brown-Henneaux formula, 
\begin{equation}
\label{BHf}
c_{2d} = \frac{3 R_{AdS_3}}{2 G_3}
\end{equation}
where the ratio between $R_{AdS_3}$ and the three dimensional Newton constant  is
\begin{equation}
\label{G3spindle}
\frac{R_{AdS_3}}{G_3} = \frac{1}{G_5} \Delta z \int_{y_N}^{y_S} e^{V(y)} | f(y) h(y) | dy.
\end{equation}
The five dimensional Newton constant for the conifold truncation and the $R_{AdS_5}$ radius are
\begin{equation}
\label{eq:G5R5}
G_5 =  \frac{8 \pi }{27 N_c^2 (g W)^3}, \quad R_{AdS_5} = g W.
\end{equation}
This can be verified by computing the 4d central charge, related to $G_5$ through the holographic relation
\begin{equation}
\label{eq:at11}
a_{T^{1,1}} = \frac{\pi  R_{AdS_5^3}}{8 G_5} = \frac{27 N_c^2}{64}
\end{equation}
where the last equality holds by plugging (\ref{eq:G5R5}) in (\ref{eq:at11}), and it corresponds to the central charge for the dual $T^{1,1}$ SCFT.
Then we must compute the integral in (\ref{BHf}). In this case we observe that 
\begin{equation}
e^{V(y)} f(y) h(y) = -\frac{\ell}{2 \kappa} (e^{3V(y)}  \cos \xi(y))'
\end{equation}
and this justifies the fact that  the central charge can be obtained only from the knowledge of the 
fields at the poles.

Furthermore in the conformal gauge the integral in the central charge becomes $e^{V(y)} |h(y)|$ and the absolute value can be removed observing as above  that
we can restrict to the region  $ h\geq  0$.
We arrive at the expression 
\begin{equation}
c_{2d} =   \frac{243 g^3 \ell N_c^2}{32 \kappa }  \sqrt{\frac{3}{2}}  (e^{3 V(y_S)} \cos \xi (y_S)-e^{3 V(y_N)} \cos \xi (y_N) ).
\end{equation}
Plugging the values of the fields evaluated at the the poles of the spindle in terms of the pole data 
the central charge becomes
 \begin{eqnarray} 
 c_{2d}= (-1)^{t_N}
\frac{3 N_c^2 (n_N+n_S) ((n_N+n_S)^2-4 p_F^2) (3(n_N+n_S)^2+4p_F^2 )}{16 \kappa  n_N n_S ((n_N^4+n_N n_S^3+n_N^3 n_S+n_S^4)-4 p_F^2 n_N n_S)}
\nonumber \\
  \end{eqnarray}
for the twist, and 
    \begin{eqnarray} 
    \label{catfin}
 c_{2d} = (-1)^{t_N+1}
\frac{3 N_c^2 (n_S-n_N) ((n_S-n_N)^2-4 p_F^2) (3(n_S-n_N)^2+4p_F^2 )}{16 \kappa  n_N n_S ((n_S^4-n_N n_S^3-n_N^3 n_S+n_N^4)-4 p_F^2 n_N n_S)}
\nonumber \\
   \end{eqnarray}
   for the antitwist.
The case of the twist is completely ruled out by this analysis because $c_{2d}>0$ is incompatible with the  requirements $M_1|_{N,S}>0$ and $e^{-12 u_3|_{N,S}} \in (0,1]$.
On the other hand the solution in the case of the anti-twist exists in the following cases:
\begin{eqnarray}
\label{variuoustN}
&&\text{for} \quad t_N=0 \,\, \& \,\,  \kappa>0 \quad \text{or} \quad t_N=1  \,\, \& \,\, \kappa<0 \quad \text{if} \quad n_N - n_S >2 |p_F| >0 \nonumber \\
&&\text{for} \quad  t_N=0  \,\, \& \,\, \kappa<0 \quad  \text{or}  \quad t_N=1  \,\, \& \,\, \kappa>0 \quad \text{if} \quad n_S- n_N >2 |p_F| >0 \nonumber \\
\end{eqnarray}

\section{Solving the BPS equations}
\label{sec:BPSsol}

\subsection{Analytic solution for the R-symmetry anti-twist}

Here we study the solutions of the BPS equations for the case of the anti-twist by turning off the 
charge for the baryonic symmetry $p_F$.
From the supergravity side this implies a further truncation to the massless graviton sector. Such a truncation always exists for a five dimensional Sasaki-Einstein manifold \cite{Buchel:2006gb,Gauntlett:2007ma}.

The truncation requires to fix the field $u_1$ to its AdS$_5$ vacuum, i.e. $u_1=0$ when fixing  $Q=4$ (see (\ref{5dvacuum})). Furthermore the analysis at the poles shows that in this case the other two scalars $u_{2,3}$ are both set to zero, compatibly with (\ref{5dvacuum}) for $Q=4$.
Observe also that this is in contrast with the assumption that the scalar $u_1$ is non vanishing in order to have a solution of the BPS equations, but it is the case only for the analytic solution that corresponds to the universal one, i.e. a further truncation to minimal gauged supergravity.
We will see that the other BPS equations can be then analitically solved in the case of the anti-twist by also fixing $p_F=0$, corresponding to the universal R-symmetry anti-twist.

The metric and the gauge fields are
\begin{eqnarray}
&&
ds^2 = 
\frac{1}{ g^2 W ^2} \left(
\frac{4 y}{9}
ds^2_{AdS_3}
+
\frac{y}{q(y)}
dy^2
+
\frac{c_0^2 q(y)}{36 y^2}
dz^2
\right),
\nonumber \\
&&
A^{(0)}= A^{(1)}= A^{(2)}
=
-\frac{1}{12} \left(\frac{c_0 \kappa}{4 g} \left(1-\frac{a}{y}\right) +\frac{s}{g}\right)
dz
\end{eqnarray}
and the function $\xi(y)$ can be expressed in terms of $q(y)$ by the relations
\begin{eqnarray}
\sin \xi=-\frac{\sqrt{q(y)}}{2 y^{3/2}},\quad
\cos \xi = \frac{\kappa  (3 y-a)}{2 y^{3/2}}
\end{eqnarray}
where $q(y)$ is 
\begin{eqnarray}
q(y) = 4 y^3-9 y^2+6 a y-a^2.
\end{eqnarray}
The constants $a$ and $c_0$ can be read from the analysis of the BPS equations from the pole data discussed above and they are
\begin{eqnarray}
a &=& \frac{\left(2 n_N+n_S\right){}^2 \left(n_N n_S+n_N^2-2 n_S^2\right){}^2}{4 \left(n_N n_S+n_N^2+n_S^2\right){}^3}, \nonumber \\
c_0&=&\frac{2 \left(n_N n_S+n_N^2+n_S^2\right)}{3 n_N n_S \left(n_N+n_S\right)}.
\end{eqnarray}
Then taking $n_S > n_N$ the  poles $y_N$ and $y_S$ are
\begin{eqnarray}
y_N = \frac{(n_S-n_N)^2 (2 n_N+n_S)^2}{4 (n_N^2+n_N n_S+n_S^2)^2}, \quad
y_S = \frac{(n_S-n_N)^2 (n_N+2 n_S)^2}{4 (n_N^2+n_N n_S+n_S^2)^2}
\end{eqnarray}
and they correspond to the two lowest roots of the polynomial $q(y)$.

Armed with these results we can compute the central charge by evaluating the integral (\ref{G3spindle}) between the two poles $y_{N,S}$ reproducing the central charge (\ref{catfin}) obtained from the  pole
data as in (\ref{variuoustN}) and by setting $p_F=0$.

\subsection{Numerical solution}
\label{numal}

The solution found by turning off $p_F$ for the anti-twist is a consistency check of the analysis, because in this case we are truncating 
to  minimal gauged supergravity, where a solution is expected \cite{Ferrero:2020laf}.
The analysis of the BPS equations from the pole data in the conformal gauge suggested the existence of a more general solution for non vanishing $p_F$.
Indeed the central charge (\ref{catfin}) is positive for suitable choices of the fluxes in the anti-twist class. Here we want to find this solution numerically for 
various numbers of $n_S, n_N$ and $p_F$ in the case of the anti-twist in the conformal gauge.

The solution is constructed by solving the BPS equation by fixing the initial conditions at one pole, for example at $y=y_N$, for $u_{2,3}$ and $V$.
Such conditions can be read from the analysis at the pole data, that also sets the value of $\ell$ necessary to find the profile for the $h(y)$ function. 
At this pole we further have $\sin \xi =0$ by assumption. On the other hand 
the initial value of $u_1$ is unfixed and indeed finding its initial value at $y=y_N$ is the task of the analysis. 

By ranging over various choices of $u_1(y_N)$ indeed the numerical solutions must interpolate the values of the other fields from $y_N$ to $y_S$.
Finding the correct and unique value of $u_1(y_N)$, up to a numerical approximation, leads then to the finite value of $y=y_S$ for which $\sin \xi=0$, 
recovering the compact geometry of the spindle.

In the following  we list a series of values of $n_S$, $n_N$ and $p_F$ for which we have obtained a solution.
In each case we have extracted the boundary values of the hyperscalar $u_1$.
Furthermore, from the numerical analysis, we have extracted also the location of the pole $y_S$, by fixing $y_N=0$.
\begin{equation}
\label{eqdata}
\begin{array}{|c|c|c|c|c|c|}
\hline
 n_S & n_N & p_F & u_1(y_N) & u_1(y_S) & y_S-y_N \\
 \hline
 7 & 1 & 1 & -0.0028068 & 0.0256628 & 2.38909 \\ \hline
 9 & 1 & 2 & 0.0823554 & -0.207454 & 2.64148 \\ \hline
  11 & 1 & 1 & 0.018792 & -0.0391945 & 2.7425 \\ \hline
   11 & 1 & 3 & 0.078249 & -0.12637 & 2.76979 \\ \hline
 9 & 3 & 1 & 0.00787918 & 0.0103843 & 1.85917 \\ \hline
 11 & 5 & 1 & 0.00866858 & 0.00952543 & 1.72325 \\ \hline
 13 & 1 & 2 & 0.0254087 & -0.0471502 & 2.8718 \\ \hline
 13 & 1 & 4 & 0.0727729 & -0.0763147 & 2.90247 \\ \hline
 13 & 3 & 1 & 0.00196656 & 0.00491044 & 2.07175 \\ \hline
 13 & 3 & 3 & 0.0219448 & 0.0352798 & 2.04444 \\ \hline
\end{array}
\end{equation}
Observe that $n_S-n_N$ is even as discussed above and we required also that $p_F$ and $\frac{1}{2}(n_S-n_N)$ have the same parity and that $2p_F<n_S-n_N$.

\begin{landscape}
\begin{figure}
\includegraphics[width=22cm]{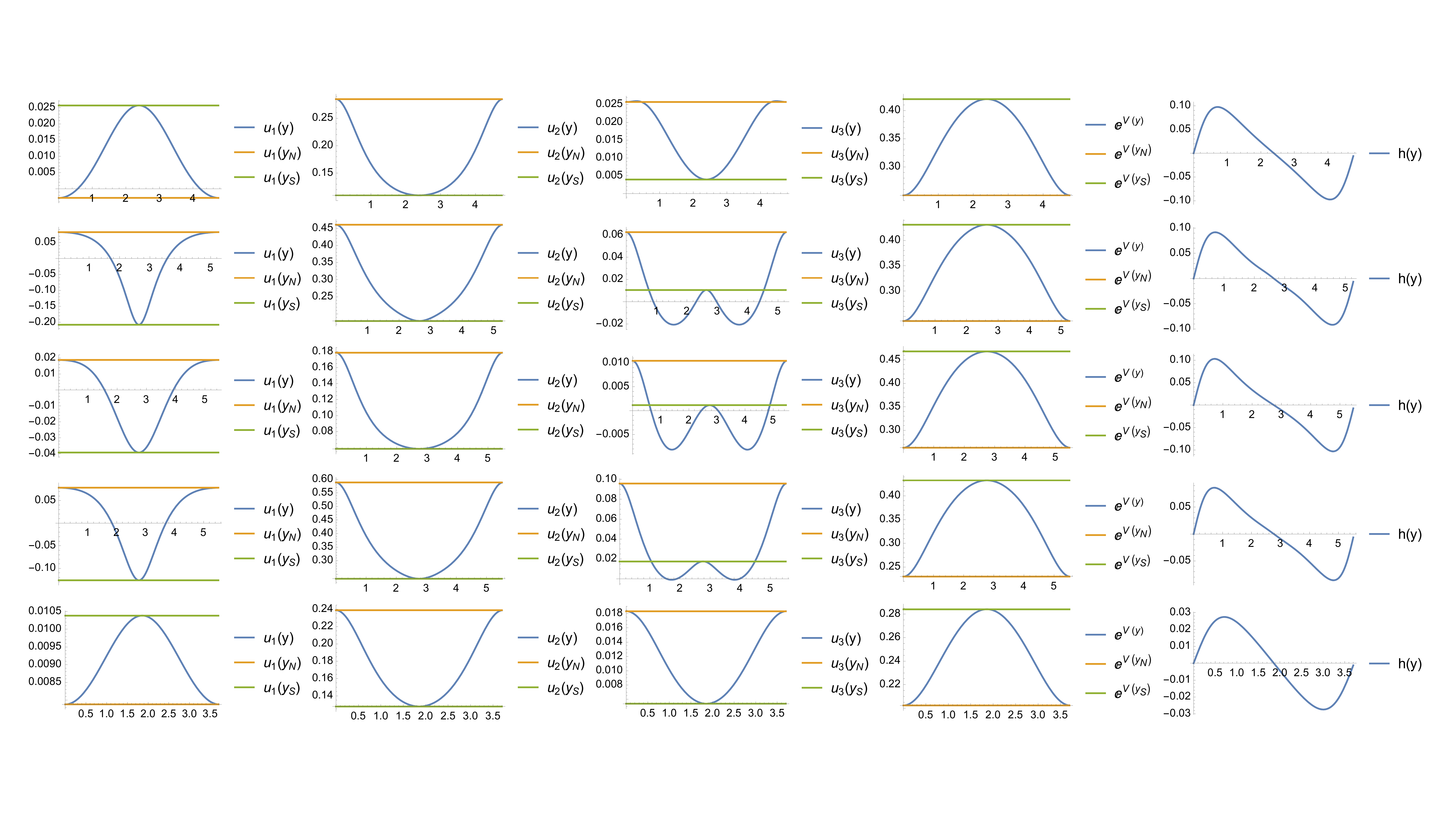}
\caption{Numerical solutions for the scalar fields $u_{1,2,3}(y)$ and the scalar functions $e^V(y)$ and $h(y)$
interpolating between the north pole at $y=y_N = 0$ and $y=2(y_S-y_N)$, where $y_S$ is the south pole..
Each line in the plot is associated to one of the first five lines in the table (\ref{eqdata}), where the values of  $n_S, n_N$  and $p_F$ are specified. These values are explicitly  $(n_S,n_N,p_F)=(7,1,1)$ for the first line,  $(9,1,2)$ for the second line, $(11,1,1)$ for the third line,  $(11,1,3)$ for the fourth line,  and $(9,3,1)$ for the fifth line.}
\label{plots1}
\end{figure}
 \end{landscape}

\begin{landscape}
\begin{figure}
\includegraphics[width=22cm]{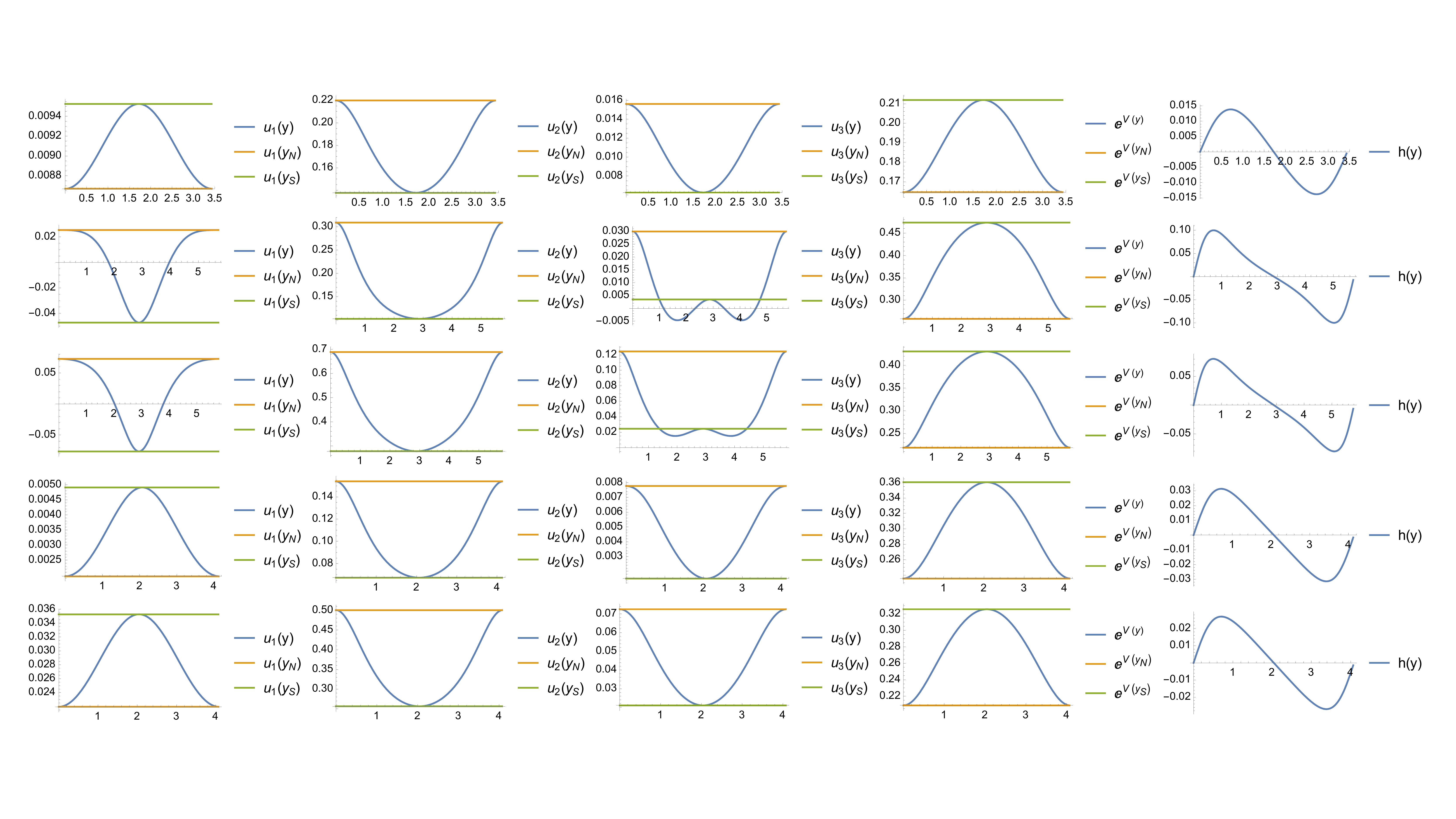}
\caption{Numerical solutions for the scalar fields $u_{1,2,3}(y)$ and the scalar functions $e^V(y)$ and $h(y)$
interpolating between the north pole at $y=y_N = 0$ and $y=2(y_S-y_N)$, where $y_S$ is the south pole..
Each line in the plot is associated to one of the last five lines in the table (\ref{eqdata}), where the values of  $n_S, n_N$  and $p_F$ are specified. These values are explicitly  $(n_S,n_N,p_F)=(11,5,1)$ for the first line,  $(13,1,2)$ for the second line, $(13,1,4)$ for the third line,  $(13,3,1)$ for the fourth line,  and $(13,3,3)$ for the fifth line.}
\label{plots2}
\end{figure}
 \end{landscape}

The explicit numerical solutions for the functions $u_{1,2,3}$, $e^V(y)$ and  $h(y)= \ell  e^{V(y)} \sin \xi(y)$ for such values are given in Figure \ref{plots1} and \ref{plots2}.
In these figures we have depicted the solutions from the north pole $y_N$ of the spindle to the south pole  $y_S$ and then we have continued the integration until $y=2(y_S-y_N)$.
In this way we have shown explicitly a consistency check of these solutions, indeed once the south pole is reached the equations reach the boundary condition that allows us to find the solutions from $y_S$ to $y_N$.

\newpage
 
\newpage

\newpage

\section{Comparison with the field theory results}
\label{sec:FT}

In this section we compare the results found above for the central charge in the anti-twist class
for the conifold truncation with respect to the calculation performed in the dual field theory.

Such calculation has been performed originally in \cite{Hosseini:2021fge} by integrating the anomaly polynomial over the geometry of the spindle. Such integration has been pursued thanks to the observation that the anomaly polynomial can be written as gluing formula in terms of the four dimensional conformal anomaly, formally expressed in terms of the R-charges, the deficit angles and the flavor fluxes.
By c-extremization the final expression, in terms of the pole data and of the quantized fluxes, is \cite{Boido:2022mbe}
\begin{equation}
\label{c2dganal}
c_{2d} =
\frac{3(m_-+\sigma m_+)^2 \sum_{a<b,c\neq a,b} \mathbf{p}_a \mathbf{p}_b \mathbf{p}_c^2 \cdot \sum_{a<b<c} \mathbf{p}_a \mathbf{p}_b \mathbf{p}_c}
{(m_-^2-\sigma m_- m_+ + m_+^2) \prod_{a<b} (\mathbf{p}_a + \mathbf{p}_b)-\sigma m_+ m_- \Theta_{KW}} N^2
\end{equation}
with
\begin{equation}
 \Theta_{KW}= \sum_{a<b,c\neq a,b} \mathbf{p}_a \mathbf{p}_b \mathbf{p}_c^4 - 2 \sum_{a<b} \mathbf{p}_a \mathbf{p}_b \prod_c \mathbf{p}_c.
\end{equation}
The comparison with the gravitational calculation requires a further restriction on the values of the fluxes. 
The fluxes $\mathbf{p}_{1,2,3,4}$ are associated to the Cartan of the global $U(1)_r \times SU(2)_L \times SU(2)_R \times U(1)_B$ of the conifold.
They are constrained by the relation $\mathbf{p}_1+\mathbf{p}_2+\mathbf{p}_3+\mathbf{p}_4  = -\frac{m_-+\sigma m_+}{m_- m_+}$
ensuring the correct quantization for the magnetic flux associated to the R-symmetry for the twist $\sigma=1$ and the anti-twist $\sigma=-1$.
The baryonic flux is instead associated to the combination $\mathbf{p}_1-\mathbf{p}_2+\mathbf{p}_3-\mathbf{p}_4$. 
The other global symmetries associated to the flavor symmetries are set to zero in the 5d supergravtiy model, that is indeed realized by truncating over the Reeb vector.
It follows that  the actual comparison requires to fix the fluxes as 
\begin{equation}
\mathbf{p}_1 = \mathbf{p}_3 = -\frac{m_-+\sigma m_+}{4m_- m_+}
+\frac{ p_b}{2 m_- m_+} 
\end{equation}
and
\begin{equation}
\mathbf{p}_2 = \mathbf{p}_4 = -\frac{m_-+\sigma m_+}{4m_- m_+} 
-\frac{ p_b}{2 m_- m_+}.
\end{equation}
Furthermore we choose $\sigma=-1$ because we have to consider the anti-twist.
Eventually we fix the pole data in the notation of \cite{Boido:2022mbe} and the ones here by
identifying $m_+ = n_N$ and $m_- = n_S$. The central charge in formula (\ref{c2dganal}) then becomes 
\begin{equation}
c_{2d}=\frac{3 N_c^2 (n_N-n_S) (3 (n_S-n_N)^2+p_b^2) ((n_S-n_N)^2-p_b^2)}{16 n_N n_S (n_N^4-n_N n_S^3-n_N^3 n_S+n_S^4-p_b^2 n_N n_S)}.
\end{equation}
This expression matches (\ref{catfin}) upon the identification $ p_F =  p_b$. 

Let us conclude this section with an observation related to the relation between our results and the general discussion that appeared in \cite{Boido:2022mbe} regarding the geometries of Type IIB constructed from a 5d SE and a spindle.
It has been observed that such geometries are constructed by fixing only the fluxes associated to the flavor symmetries, i.e. one can always turn off the baryonic symmetry to obtain the internal Gauntlett-Kim 7d geometry \cite{Gauntlett:2007ts}.
We expect that the solutions found here,  when uplifted to 10d are in the  class of \cite{Gauntlett:2007ts}, with just five-form flux turned on.

\section{Conclusions}
\label{sec:conc}

In this paper we studied a supersymmetric AdS$_3 \times \Sigma$ asymptotic to the AdS$_5$ $\mathcal{N}=2$ truncation of the conifold with a Betti vector multiplet
found in \cite{Cassani:2010na}.
The model consists of gauged supergravity with two vector multiplets and two hypermultiplets. The vector fields gauge a subgroup of the quaternionic manifold and 
one gauge field becomes massive via Higgs mechanism.
In the low energy spectrum there are then two massless fields, the graviphoton and the Betti vector.
One is associated to the R-symmetry and the other one to the baryonic symmetry of the dual Klebanov-Witten field theory.
When this model is compactified on the spindle many of the scalars in the hypermultiplet can be further truncated.
A crucial aspect of this compactification is that we need to include some of the scalars from the hypermultiplet in the analysis.
After a suitable ansatz on the dependence of the spindle coordinates from the scalar fields, we have computed the BPS and the 
gauge fields equations.  This has allowed us to find the properly quantized fluxes through the spindle.
We have shown that the analysis of the fluxes can be performed by studying the BPS equations  at the poles of the spindle 
and that this analysis, in the conformal gauge, fixes the proper boundary conditions for all the scalars except the ones in the hypermultiplet.
Furthermore thanks to this analysis it has been possible to compute the central charge of the would be AdS$_3$ solution from the Brown-Henneaux formula.
By inspection we have observed that only the anti-twist class is consistent with the unitarity bound requiring a positive central charge.
This analysis is however not enough to claim the existence of the AdS$_3$ solution and for this reason it is necessary to solve the BPS equations explicitly 
from the north to the south pole of the spindle.
By turning off the baryonic magnetic charge we have observed that the universal solution of \cite{Ferrero:2020laf} is recovered. Furthermore we have provided a numerical analysis for the 
case with the magnetic baryonic charge turned on. In the numerical analysis we have first imposed the boundary conditions obtained from the analysis of the BPS equations at the north pole. 
Then we have scanned  the boundary value of the scalar in the hypermultiplet that we cannot turn off. Solving the BPS equations for various initial data we have looked for the unique 
solution corresponding to the existence of a compact spindle. Once this value has been found we have checked that indeed the other fields take the value expected from the 
analysis of the pole data at the south pole, where this last value has been extracted numerically.
We have eventually  compared  our general expression of central charge for the anti-twist  with the one obtained in \cite{Hosseini:2021fge,Boido:2022mbe} from the dual  field theory, 
restricted to the baryonic anti-twist, and we have found an exact agreement.

Many interesting  directions in the study of supergravity truncations on the spindle deserve further investigation.
For example one can study other consistent 5d $\mathcal{N}=2$ truncations with vector multiplets and hypermultiplets where an holographic dual field theory is available. From the analysis performed here and previously in \cite{Boido:2022mbe} one can observe that the central charge can be extracted from the pole data if the number  of conserved charges $n_q$  is twice the number of massless vector multiplets  $n_v^{m^2=0}$. 
We observe that  $n_v^{m^2=0}=1$ because the number of conserved charges corresponds to $n_v^{m^2=0}+1$.
If this is not the case then one should provide a more general solution of the BPS equations from the numerical analysis. The problem in this last case is that the boundary conditions for some of the scalars in the vector multiplets must be fixed numerically together with the ones for the scalar(s) in the hypermultiplet. Searching for such initial data is in general non-trivial and it may require a refinement of the numerical techniques. An interesting truncation that has  $n_v^{m^2=0}=1$ is the one recently worked out in \cite{Cassani:2020cod} elaborating on previous results \cite{Szepietowski:2012tb}. This truncation describes a stack of M5 branes  wrapped on a two dimensional Riemann surface $\Sigma$. In this case the holographic dual field theories have been found in \cite{Bah:2011vv,Bah:2012dg}. 
We expect that the compactification of this truncation on the spindle  proceeds along the same lines of the discussion performed here and in \cite{Arav:2022lzo} and that the central charge in this case can be extracted from the pole data as well. It should be interesting to study this case  by turning on a magnetic charge for the flavor symmetry in order to obtain the solution of the BPS equations interpolating between the poles of the compact spindle. We are currently investigating in this direction.

Another interesting aspect discussed in \cite{Arav:2022lzo} consists of reformulating the BPS equations in a D=4 Janus form in order to interpret the conserved charges from a different perspective. We expect that similar results can be provided for the model discussed in this paper as well.

A more complicated question regards the interpretation of the solutions found here from a flow across dimensions along the radial direction. This requires a modification of the anstaz (\ref{ads3ansatz}) and requires to solve the BPS equations also for the radial profiles of the scalars.
Finding such solutions is a necessary step to obtain a supergravity attractor mechanism dual to  
c-extremization on the spindle.

We conclude by observing that one can also look at truncations dual to 3d SCFT, corresponding to AdS$_4$  $\mathcal{N}=2$ gauged supergravity in presence of hypermultiplets.
In this case an attempt has  recently in \cite{Suh:2022pkg} where the case of the mABJM model has been considered.
In this case one has $n_v^{m^2=0}=2$ and the counting discussed above implies that the analysis at the poles does not entirely fixe the boundary conditions for the scalar fields in the vector multiplets.
Similar results are expected for another truncation with a known dual field theory,  corresponding to the case of massive Type IIA supergravity \cite{Guarino:2015jca}.

%%%%%%%%%%%%%%%%%%
%
%
%
%
%
%
%%%%%%%%%%%%%%%%%%
\section*{Acknowledgments}
%%%%%%%%%%%%%%%%%%
%
%
We are grateful to Jerome Gauntlett useful comments on the manuscript. We further acknowledge 
Minwoo Suh for informing of \cite{Suh:2023xse}.
The work of A.A. and A.S. has been supported in part by the Italian Ministero dell'Istruzione, 
Universit\`a e Ricerca (MIUR), in part by Istituto Nazionale di Fisica Nucleare (INFN) through the “Gauge Theories, Strings, Supergravity” (GSS) research project and in part by MIUR-PRIN contract 2017CC72MK-003. The work of N.P. is supported by the Israel Science Foundation (grant No. 741/20) and by the German Research Foundation through a German-Israeli Project Cooperation (DIP) grant ``Holography and the Swampland".

\appendix

\section{Matter-coupled $\ma N=2$ 5d gauged supergravity}\label{generalSUGRA}

In this appendix we summarize the general features of matter-coupled $\ma N=2$ gauged supergravities in five dimensions following the conventions of \cite{Bergshoeff:2002qk,Bergshoeff:2004kh}. The $\ma N=2$ 5d supergravity multiplet can be coupled to $n_V$ vector multiplets and $n_H$ hypermultiplets. Restricting us only to the bosonic sector, the field content of this class of theories is determined by the gravitational field $g_{\mu\nu}$, $n_V$ real moduli $\phi^x$ belonging to vector multiples with $x=1,\dots, n_V$, $4n_H$ real scalars (or hyperscalars) $q^X$ from the hypermultiplets with $X=1,\dots, 4n_H$ and $n_V+1$ vector fields $A^I_\mu=(A_\mu^0, A_\mu^x)$ with $I=(0, x)$. The vector $A_\mu^0$ belongs to the supergravity multiplet and it is usually called the graviphoton; the vectors $A_\mu^x$ are included in the vector multiplets.

Let's briefly discuss the properties of the $\ma N=2$ moduli space.
The scalars $\phi^x$ parametrize a $n_V$-dimensional Very Special Real manifold. This scalar geometry can be defined firstly introducing a $(n_V+1)$-dimensional embedding manifold parameterized by the homogeneous coordinates $h^I(\phi^x)$. Secondly considering the hypersurface identified by the fundamental cubic polynomial,
\begin{equation}
 P(h)=C_{IJK}h^Ih^Jh^K=1\,,
\end{equation}
where $C_{IJK}$ is a constant and symmetric tensor. From the tensor $C_{IJK}$ and the homogeneous coordinates $h^I$ is thus possible to derive all the quantities needed to describe the couplings to vector multiplets. In fact on such hypersurface we can introduce a metric $g_{xy}$ as the pull-back of the metric $a_{IJ}$ on the embedding space,
\begin{equation}
 g_{xy}= \frac{3}{2} \partial_xh^I\partial_yh^J\,a_{IJ}\qquad \text{with}\qquad a_{IJ}=-\frac13\,\partial_{h^I}\partial_{h^J}\,\log P(h)|_{P=1}\,.
\end{equation}
The coupling between scalars $\phi^x$ and vectors $A^I_\mu$ is realized through the metric $a_{IJ}$ which can be alternatively defined as
\begin{equation}\label{SKa}
 a_{IJ}=-2C_{IJK}h^K+3h_Ih_J     \qquad \text{with}\qquad h_I \equiv a_{IJ}h^J=C_{IJK}h^Jh^K\,.
\end{equation}

The hyperscalars $q^X$ parametrize a Quaternionic manifold which can be defined as a Riemannian manifold with metric $g_{XY}$ endowed with a locally defined triplet $\vec J_{XY}$ of almost complex structures satisfying the relations
\begin{equation}
\begin{split}\label{quaternionicrelations}
& g^{ZW}\,J^r_{XZ}J^s_{WY}=-\delta^{rs}g_{XY}+\varepsilon^{rst}J^t_{XY}\,,\\
& \widetilde{\nabla}_Z \vec{J}_X^{\ \, Y}=\nabla_Z \vec J_X^{\ \,Y}+2 \, \vec \omega_Z\times \vec J_X^{\ \,Y}=0
 \end{split}
\end{equation}
where $\{r, s, t\}=1,2,3$ are indices of the vectorial representation of $SU(2)$\footnote{In this paper we use also the doublet notation instead of the vector (or triplet) one,
  \begin{equation}
    \label{DoubletToTriplet}
    J_{X \ \ i}^{\ \ Y  \ j} \equiv i \,\vec J_X^{\ \, Y} \cdot \vec{\sigma}_i^{\ j},
  \end{equation}
  where $\vec{\sigma}_i^{\ j}$ are the three Pauli matrices. Of course, this transition between triplet and doublet notation holds also for other quantities in the adjoint representation of $SU(2)$.}. The first expression in \eqref{quaternionicrelations} is the defining relation of a quaternionic structure. The second implies that $\vec J$ is covariantly constant w.r.t. the connection $\vec \omega=\vec \omega_X(q)dq^X$ associated to a $SU(2)$-bundle defined over the quaternionic manifold. The corresponding $SU(2)$-curvature $\vec{\ma R}$ turns out to be proportional to the complex structures $\vec J_{XY}$ as it follows
\begin{equation}
  \label{mR:J}
 \vec{\ma R}=d\vec\omega+2\, \vec\omega\wedge \vec\omega=\frac12\, \nu \vec J\, \qquad \text{with} \qquad \nu \equiv \frac{1}{4 n_H (n_H + 2)} \, R.
\end{equation}

In this paper we focus only on abelian gaugings on the quaternionic manifold. These are typically generated by commuting Killing vectors $k_I^X(q)$. For each Killing vector we can introduce a triplet of moment maps $\vec P_I$ through the relation
\begin{equation}
\begin{split} \label{equivariance}
 &\widetilde{\nabla}_X\vec P_I=\partial_X\vec P_I+2\, \vec\omega_X\times \vec P_I= \vec J_{XY}k^Y_I\,.\\
 %&\vec P_I\times \vec P_J=-\frac{1}{2}\,\vec J_{XY}k_I^Xk_J^Y\,.
 \end{split}
\end{equation}
We are now ready to write the bosonic Lagrangian,
\begin{equation}
  \label{L:5d}
  \begin{split}
    e^{-1} \mathscr{L} = & \,\,\frac12R - \frac{1}{2} g_{xy}(\phi) \partial_\mu \phi^x \partial^\mu \phi^y - \frac{1}{2} g_{XY}(q) \mathcal{D}_\mu q^X \mathcal{D}^\mu q^Y - \frac{1}{4} a_{IJ} (\phi) F^I_{\mu\nu} F^{J \, \mu\nu} \\
    & + \frac{e^{-1}}{12}\sqrt{\frac{2}{3}} C_{IJK} \epsilon^{\mu\nu\rho\sigma\tau} F^I_{\mu\nu} F^J_{\rho\sigma} A^K_\tau - g^2 V(\phi,q),
  \end{split}
\end{equation}
where 
\begin{equation}
 \mathcal{D}_\mu q^u = \partial_\mu q^u + g  A^I_\mu k^u_I
 \end{equation}
 are the covariant derivatives of hyperscalars depending on the gauge coupling $g$. Finally the scalar potential $V(\phi,q)$ has the form
\begin{equation}
  \begin{split}\label{scalarpotential}
    V =% & \, 4 \vec{P} \cdot \vec{P} - 2 \vec{P}^x \cdot \vec{P}_x - 2 \mathcal{N}_{iA} \mathcal{N}^{iA} \\
     & P_I^r P_J^r \Bigl( 4 h^I h^J - 3 g^{xy} \partial_x h^I \partial_y h^J \Bigr) - \frac{3}{4} g_{XY} k_I^X k_J^Y h^I h^J\,.
  \end{split}
\end{equation}
We point out that since the gauging is abelian, the coviarant derivatives of the scalars $\phi^x$ are just the ordinary onces since the scalars $\phi^x$ transform trivially under abelian symmetries.

Finally we can write the $\ma N=2$ supersymmetry variations using the conventions of \cite{Bergshoeff:2004kh, Halmagyi:2011yd, Louis:2016msm}. Given a $\ma N=2$ Killing spinor $\epsilon^i$, the fermionic variations have the following form
\begin{align}
  \delta \psi_\mu^i & = \bigl[ D_\mu + \frac{i}{4 \sqrt{6}} h_I (\gamma_\mu^{\ \nu \rho} -4 \delta_\mu^\nu \gamma^\rho) F^I_{\nu\rho} \bigr] \epsilon^i - \frac{i}{\sqrt{6}} g \, \gamma_\mu h^I (P_I)^{i j} \epsilon_j, \label{susy:gr} \\ 
  \delta \lambda^{x \, i} & = \bigl( -\frac{i}{2} \gamma^\mu \partial_\mu \phi^x + \frac{1}{4}  \gamma^{\mu\nu} F^I_{\mu\nu} h^x_I \bigr) \epsilon^i - g \,P^{x \, i j} \epsilon_j, \label{susy:g} \\
  \delta \zeta^A & = \frac{i}{2} \gamma^\mu \mathcal{D}_\mu q^X f^{iA}_X \epsilon_i - \frac{\sqrt{6}\,g}{4}h^I\,k_I^X f^{i A}_{X} \epsilon_i, \label{susy:h}
\end{align}
where we have introduced the quaternionic vielbein $f^{i A}_X$ \cite{Bergshoeff:2004kh} and the supercovariant derivative
\begin{equation}
\begin{split} 
 &  D_\mu \epsilon^i = \partial_\mu \epsilon^i + \frac{1}{4} \omega^{ab}_\mu \gamma_{ab} \epsilon^i - \partial_\mu q^X \omega^{ij}_X \epsilon_j - g  A_\mu^I P_I^{i j} \epsilon_j\,,\\
 %&\ma N_i^A=\frac{\sqrt{6}}{4}h^I\,k_I^X f^{i\,A}_X
   \end{split}
\end{equation}
acting on the supersymmetry parameter $\epsilon^i$.\\
The vielbein $f^{i A}_X$ satisfies the following relation with the almost complex structures:
\begin{equation}
  \label{Jtof}
  i \, \vec{J}_X^{\ \, Y} \cdot \vec{\sigma}_i^{\ j} = 2 f_X^{j  A} f^Y_{i  A} - \delta_i^j \delta_X^Y,
\end{equation}
that will be useful in the simplification of the BPS equations.

\section{Details on the quaternionic geometry}
\label{sec:app1}
%{\it citare da qualche parte anche  \cite{DallAgata:2021nnr}}\\
In this appendix we derive the quaternionic structures $\vec J$ and the $SU(2)$ spin connections $\vec \omega$ for the quaternionic manifold 
\begin{equation}
  \frac{SO(4,2)}{SO(4) \times SO(2)}.
\end{equation}
This scalar geometry was already studied in detail in \cite{Halmagyi:2011yd, Louis:2016msm,DallAgata:2021nnr} and it is described by $n_H=2$ hypermultiplets, whose real scalars can be organized as $q^X, \ X=1, \dots, 8$%, parametrize the quaternionic-K\"ahler geometry of the coset space
%\begin{equation}
 % \frac{SO(4,2)}{SO(4) \times SO(2)}.
%\end{equation}

Following \cite{Halmagyi:2011yd, Louis:2016msm} we use the coordinates $q^X = \{u_1, k, a, \phi, b^1 ,\bar{b}^1, b^2, \bar{b}^2 \}$, with line element given by
\begin{equation}
  \begin{split}
    g_{XY}  dq^X dq^Y = & -2 e^{-4u_1} M_{ij} \bigl( b^i d\bar{b}^j + \bar{b}^i d b^j \bigr) - 4 du_1^2  - \frac{1}{4} d\phi^2 - \frac{1}{4} e^{2\phi} da^2  \\
    & - \frac{1}{4} e^{-8u_1} \Bigl[ dk + 2 \varepsilon_{ij} \bigl( b^i d\bar{b}^j + \bar{b}^i db^j  \bigr)  \Bigr]^2 ,
  \end{split}  
\end{equation}
where
\begin{equation}
  M_{ij} = e^\phi 
  \begin{pmatrix}
    a^2 + e^{-2\phi} & \quad -a \\
    -a & 1
  \end{pmatrix}
  .
\end{equation}
We have chosen the normalization of the metric in order to have the curvature scalar $R=64$ and thus $\vec{\mathcal{R}} = \vec{J}$ (see \eqref{mR:J}). This last equality is necessary to obtain the simplified version of the hyperino variation in \eqref{hyp:var:2}.

To display the quaternionic structure of the coset, we follow the explicit construction made in Appendix E of \cite{Cheung:2021igt}\footnote{See also  \cite{DallAgata:2021nnr} for a similar analysis.}, adapting it to our coordinates, and we introduce the quaternionic vielbeins
\begin{equation}
  \begin{split}
    f^1 & = \frac{i}{2} d\phi, \qquad f^2 = 2 i \, du_1, \qquad f^3 = \frac{i}{2} e^\phi da, \\
    f^4 & = i e^{-4u_1} \biggl( \frac{1}{2} dk -  \bar{b}^2 db^1 +  \bar{b}^1 db^2 - b^2 d\bar{b}^1 + b^1 d\bar{b}^2 \biggr), \\
    f^5 & = \frac{i}{\sqrt{2}} e^{-2u_1-\frac{\phi}{2}} \bigl( db^1+d\bar{b}^1 \bigr), \qquad  f^6 = \frac{1}{\sqrt{2}} e^{-2u_1-\frac{\phi}{2}} \bigl( db^1-d\bar{b}^1 \bigr),\\
    f^7 & = \frac{i}{\sqrt{2}} e^{-2u_1+\frac{\phi}{2}} \bigl( -a \, db^1 + db^2 - a \, d\bar{b}^1 + d\bar{b}^2 \bigr),\\
    f^8 & = -\frac{1}{\sqrt{2}} e^{-2u_1+\frac{\phi}{2}} \bigl( a \, db^1 - db^2 + a \, d\bar{b}^1 - d\bar{b}^2 \bigr).
  \end{split}
\end{equation}
We can thus obtain the triplet of almost complex structures $\vec{J}$ as
\begin{equation}
  \begin{split}
    J^1 & = \frac{1}{\sqrt{2}} \Bigl( f^{15} + f^{18} + f^{25} - f^{28} - f^{36} + f^{37} - f^{46} - f^{47} \Bigr),\\
    J^2 & = \frac{1}{\sqrt{2}} \Bigl( f^{16} - f^{17} + f^{26} + f^{27} + f^{35} + f^{38} + f^{45} - f^{48} \Bigr),\\
    J^3 & = - \Bigl( f^{13} + f^{24} + f^{56} + f^{78} \Bigr),
  \end{split}
\end{equation}
where $f^{ij} \equiv f^i \wedge f^j$ and $J^r = \frac{1}{2} J^r_{\ XY} \, dq^X \wedge dq^Y$ for $r=1,2,3$. One can check that these structures satisfy the quaternionic relations in \eqref{quaternionicrelations}.
In this setting, the $SU(2)$ connections take the form
\begin{equation}
  \begin{split}
    \omega^1 & = \frac{1}{2}  e^{-2u_1 - \frac{\phi}{2}} ( db^1 + d \bar{b}^1 )  - \frac{i}{2} a \, e^{-2u_1 + \frac{\phi}{2}}( d b^1 - d\bar{b}^1 ) + \frac{i}{2} e^{-2u_1 + \frac{\phi}{2}} ( db^2 - d\bar{b}^2 ),\\
    \omega^2 & = - \frac{i}{2} e^{-2u_1 - \frac{\phi}{2}} ( db^1 - d \bar{b}^1 )  - \frac{1}{2} a \, e^{-2u_1 + \frac{\phi}{2}}( d\bar{b}^1 + d\bar{b}^1 ) + \frac{1}{2} e^{-2u_1 + \frac{\phi}{2}} ( db^2 + d\bar{b}^2 ),\\
    \omega^3 & = \frac{1}{4} e^\phi da -\frac{1}{2} e^{-4u_1} \biggl( \frac{1}{2} dk -\bar{b}^2 db^1 + \bar{b}^1 db^2 - b^2 d\bar{b}^1 +b^1 d\bar{b}^2  \biggr).
  \end{split}
\end{equation}

\section{Derivation of the BPS equations}\label{app:BPS}

Following the analysis made in \cite{Arav:2022lzo}, in order to construct AdS$_3 \times \Sigma$ solutions to the BPS equations, we first decompose the Clifford algebra via
\begin{equation}
  \gamma^m = \Gamma^m \otimes \sigma^3, \qquad \gamma^3 = \mathbb{I}_2 \otimes \sigma^1, \qquad \gamma^4 = \mathbb{I}_2 \otimes \sigma^2,
\end{equation}
where $\Gamma^m = \bigl( -i \sigma^2, \sigma^3, \sigma^1 \bigr)$\footnote{We are using the mostly plus signature, while in \cite{Arav:2022lzo} they use the mostly minus one.} are the gamma matrices in $d=3$ and $\sigma^i, \ i=1,2,3,$ are the Pauli matrices. We can thus write the Killing spinor as
\begin{equation}
  \epsilon = \psi \otimes \chi,
\end{equation}
where $\chi$ is a two-component spinor on the spindle and $\psi$ is a two-component spinor on AdS$_3$ satisfying
\begin{equation}
  \nabla_m \psi = - \frac{\kappa}{2} \, \Gamma_m \psi,
\end{equation}
where $\kappa = \pm 1$ specifies the chirality of the supersymmetry of the dual 2d SCFT. In this section we will analyze the BPS equations in order to determine the structure of the spinor $\chi$, that is given by
\begin{equation}
  \chi = e^{V/2} e^{i s z}
  \begin{pmatrix}
    \sin \frac{\xi}{2} \\
    \cos \frac{\xi}{2}
    \end{pmatrix}
\end{equation}
as we will see in more detail in a few.

\subsection{Gravitino variation}
The supersymmetry variation for the gravitino in \eqref{susy:gr} splits into two decoupled equations if we impose the projection condition on the symplectic-Majorana $\epsilon^i$ \footnote{See appendix A.2.1 of \cite{Baggio:2014hua} for a more general overview on the projections of $SU(2)$ symplectic-Majorana fermions.}. This decoupling is due to the fact that in our model only the $r=3$ $SU(2)$-components of the moment maps survive (see \eqref{momentMapsTruncated}), which are related to the doublet notation through the third Pauli matrix via \eqref{DoubletToTriplet}.\\  
Thus, one of the two BPS equations obtained from the gravitino variation can be written as
\begin{equation}
  \label{var:gr:2}
  \delta \psi_\mu = \Bigl[ \nabla_\mu -i Q_\mu + \frac{i}{12} H_{\nu\rho} \bigl(\gamma^{\nu\rho} \gamma_\mu + 2 \gamma^\nu \delta_\mu^\rho \bigr) + \frac{1}{2} g  W \gamma_\mu \Bigr] \epsilon = 0,
\end{equation}
where $\epsilon$ is now a Dirac spinor, $\nabla_\mu \epsilon = \partial_\mu \epsilon + \frac{1}{4} \omega_{\mu a b} \gamma^{ab}$ and  $Q_\mu \equiv \partial_\mu q^X \omega_X^3 + g  A_\mu^I P_I^3 $. We also introduced the superpotential $W \equiv \sqrt{\frac{2}{3}} h^I P_I^3$ and $H_{\mu\nu} \equiv h_I F^I_{\mu\nu}$.\\
The components of \eqref{var:gr:2} that are tangent to the directions along AdS$_3$ give:
\begin{equation}
  \label{eq:gr:1}
  \bigl[ - \bigl( \kappa e^{-V} + \frac{1}{\sqrt6} H_{34} \bigr) \gamma^{34} + i \, V' f^{-1} \gamma^3 \bigr] \epsilon = -i g W \epsilon.
\end{equation}
In order to have non-trivial solutions to this equation, we have to impose that the two coefficients on the left hand side live on a circle, i.e.
\begin{equation}
  \label{proj}
  \bigl[ \cos\xi \gamma^{34} + i \sin\xi \gamma^3 \bigr] \epsilon = - i \epsilon.
\end{equation}
This projection condition is solved by
\begin{equation}
  \epsilon = e^{-i \frac{\xi}{2} \gamma^4} \eta, \qquad \gamma^3 \eta = i \gamma^4 \eta
\end{equation}
and it allows us to split \eqref{eq:gr:1} in
\begin{equation}
  \label{eq:gr:2}
     -  \kappa e^{-V} -  \frac{1}{\sqrt6} H_{34} =  g W \cos\xi, \qquad  V' f^{-1} =  g W \sin\xi.
\end{equation}
If we now focus on the component of the gravitino variation in the longitudinal direction of the spindle, i.e. $\mu=y$, we can rewrite it as
\begin{equation}
  \label{eq:gr:y}
  \Bigl[ \partial_y -\frac{1}{2} V' - \frac{i}{2} \bigl( \partial_y \xi + \sqrt{\frac32} f H_{34} + \kappa f e^{-V} \bigr) \gamma^4 \Bigr] \eta = 0. 
\end{equation}
One can notice that this expression is in the form $(a_1 + a_2 \gamma^4 ) \eta = 0$, which implies that $a_1^2 + a_2^2 = 0$.\\
Therefore, from the first part of \eqref{eq:gr:y} we can infer that $\eta$ has the structure
\begin{equation}
  \label{eta}
  \eta = e^{V/2} e^{isz} \eta_0,
\end{equation}
where $s$ is a constant and $\eta_0$ is independent from $y$ and $z$. From the second part of \eqref{eq:gr:y} we obtain
\begin{equation}
  \label{eq:gr:y:2}
  \partial_y \xi + \sqrt{\frac32} f H_{34} + \kappa f e^{-V} = 0.
\end{equation}
Similarly, the component along the azimuthal direction of the spindle ($\mu=z$) gives
\begin{equation}
  \label{eq:gr:z}
  \begin{split}
    \Bigl[ \partial_z - i Q_z & + \frac{i}{2} f^{-1} h' \cos\xi - \frac{i}{\sqrt6} H_{34} h \sin\xi  \\
    & + \Bigl( - \frac{1}{2} f^{-1} h' \sin\xi + \frac{1}{2} g W h - \frac{1}{\sqrt6}H_{34} h \cos\xi \Bigr) \gamma^4 \Bigr] \eta = 0,
  \end{split} 
\end{equation}
from which
 \begin{align}
   & (s - Q_z) + \frac{1}{2} f^{-1} h' \cos\xi - \frac{1}{\sqrt6} H_{34} h \sin\xi = 0,\\
   & -\frac{1}{2} f^{-1} h' \sin\xi + \frac{g W h}{2} - \frac{1}{\sqrt6} H_{34} h \cos\xi = 0.
 \end{align}

 \subsection{Gaugino variation}
 Using some relations of the Very Special Real geometry and the definition of the superpotential, the variation of the gaugino \eqref{susy:g} gives
\begin{equation}
   \delta \lambda^x  = \Bigl[ -\frac{i}{2} \gamma^\mu \partial_\mu \phi^x + \frac{1}{4} \sqrt{\frac{3}{2}} g^{xy} \partial_y h_I \gamma^{\mu\nu} F^I_{\mu\nu} + i \sqrt{\frac{3}{2}} g \, g^{xy} \partial_y W \Bigr] \epsilon = 0.
 \end{equation}
 From the first component ($x=1$), imposing again the projection condition in \eqref{proj}, we obtain
 \begin{align}
   \label{BPSeqGaugino:1}
   & f^{-1} u_2' + \frac{3 g}{4} \partial_{u_2}W \sin\xi = 0,\\
   & 3 g \, \partial_{u_2}W \cos\xi + \sqrt{\frac{2}{3}} e^{2 u_3} \bigl( e^{-2u_2} F_{34}^{(1)} - e^{2u_2} F_{34}^{(2)} \bigr) = 0,
 \end{align}
 where we have used the explicit expressions for the sections $h^I$ and for the field strengths. Similarly, from the component $x=2$, we have
 \begin{align}
   \label{BPSeqGaugino:2}
   & f^{-1} u_3' + \frac{g}{4} \partial_{u_3}W \sin\xi = 0,\\
   & 3 g \, \partial_{u_3}W \cos\xi + \sqrt{\frac{2}{3}} \bigl( 2 e^{-4u_3} F_{34}^{(0)} - e^{-2u_2 + 2u_3} F_{34}^{(1)} - e^{2u_2+2u_3} F_{34}^{(2)} \bigr) = 0.
 \end{align}

 \subsection{Hyperino variation}
 In order to simplify the BPS equation coming from the hyperino variation in \eqref{susy:h}, we can multiply its expression for $f_{jAY}$. Using \eqref{Jtof}, after some calculations we obtain the relation
\begin{equation}
  \label{hyp:var:2}
  \Bigl( 2 i \gamma^\mu \partial_\mu u_1 + \frac{3}{8} i g \partial_{u_1} W + \frac{1}{4} \partial_{u_1} Q_\mu \gamma^\mu \Bigr) \epsilon = 0.
\end{equation}
Notice that to single out the vector $Q_\mu$, introduced in \eqref{var:gr:2}, it is necessary to make a precise choice of the normalization of the metric of the quaternionic manifold, as we have pointed out in appendix \ref{sec:app1}.\\
Finally, imposing the projection condition \eqref{proj}, this last equation gives
\begin{align}\label{BPSeqHyperino}
  f^{-1} u_1' & = \frac{3g}{16} \frac{\partial_{u_1} W}{\sin\xi}, \\
  \frac{3g}{2} \partial_{u_1}W \cos\xi & = h^{-1} \partial_{u_1} Q_z \sin\xi.
\end{align}

\bibliographystyle{JHEP}
\bibliography{ref}

\end{document}